\documentclass{aastex63}

\usepackage{amsmath}
\usepackage{graphicx}
\usepackage{grffile}

\submitjournal{AJ}

\received{}
\revised{}
\accepted{}

\shorttitle{2012 TC4}
\shortauthors{Lee et al.}

\graphicspath{{./}{figures/}}

\begin{document}

\title{Spin Change of Asteroid 2012 TC4 probably by Radiation Torques}

\correspondingauthor{Hee-Jae Lee}
\email{hjlee@kasi.re.kr}

\author{Hee-Jae Lee}
\affiliation{Chungbuk National University, 1 Chungdae-ro, Seowon-gu, Cheongju, Chungbuk 28644, Korea}
\affiliation{Korea Astronomy and Space Science Institute, 776, Daedeokdae-ro, Yuseong-gu, Daejeon 34055, Korea}

\author{Josef \v{D}urech}
\affiliation{Astronomical Institute, Faculty of Mathematics and Physics, Charles University, V Hole\v{s}ovi\v{c}k\'ach~2, 180 00 Prague~8, Czech Republic}

\author{David Vokrouhlick\'{y}}
\affiliation{Astronomical Institute, Faculty of Mathematics and Physics, Charles University, V Hole\v{s}ovi\v{c}k\'ach~2, 180 00 Prague~8, Czech Republic}

\author{Petr Pravec}
\affiliation{Astronomical Institute, Academy of Sciences of the Czech Republic, Fri\v{c}ova~1, 251 65 Ond\v{r}ejov, Czech Republic}

\author{Hong-Kyu Moon}
\affiliation{Korea Astronomy and Space Science Institute, 776, Daedeokdae-ro, Yuseong-gu, Daejeon 34055, Korea}

\author{William Ryan}
\affiliation{New Mexico Institute of Mining and Technology, Socorro, NM 87801, USA}

\author{Myung-Jin Kim}
\affiliation{Korea Astronomy and Space Science Institute, 776, Daedeokdae-ro, Yuseong-gu, Daejeon 34055, Korea}

\author{Chun-Hwey Kim}
\affiliation{Chungbuk National University, 1 Chungdae-ro, Seowon-gu, Cheongju, Chungbuk 28644, Korea}

\author{Young-Jun Choi}
\affiliation{Korea Astronomy and Space Science Institute, 776, Daedeokdae-ro, Yuseong-gu, Daejeon 34055, Korea}
\affiliation{University of Science and Technology, 217, Gajeong-ro, Yuseong-gu, Daejeon 34113, Korea}

\author{Paolo Bacci}
\affiliation{Gruppo Astrofili Montagna Pistoiese, 51028 San Marcello Piteglio, Province of Pistoia, Italia}

\author{Joe Pollock}
\affiliation{Physics and Astronomy Department, Appalachian State University, Boone, NC 28608, USA}

\author{Rolf Apitzsch}
\affiliation{Wildberg Observatory, 72218 Wildberg, Germany}

\begin{abstract}
 Asteroid 2012~TC4 is a small ($\sim$10\,m) near-Earth object that was observed during its Earth close approaches in 2012 and 2017. Earlier analyses of light curves revealed its excited rotation state. We collected all available photometric data from the two apparitions to reconstruct its rotation state and convex shape model. We show that light curves from 2012 and 2017 cannot be fitted with a single set of model parameters -- the rotation and precession periods are significantly different for these two data sets and they must have changed between or during the two apparitions. Nevertheless, we could fit all light curves with a dynamically self-consistent model assuming that the spin states of 2012~TC4 in 2012 and 2017 were different. To interpret our results, we developed a numerical model of its spin evolution in which we included two potentially relevant perturbations: (i) gravitational torque due to the Sun and Earth, and (ii) radiation torque known as the Yarkovsky-O'Keefe-Radzievskii-Paddack (YORP) effect. Despite our model simplicity, we found that the role of gravitational torques is negligible. Instead, we argue that the observed change of its spin state may be plausibly explained as a result of the YORP torque. To strengthen this interpretation we verify that (i) the internal energy dissipation due to material inelasticity, and (ii) an impact with a sufficiently large interplanetary particle are both highly unlikely causes its observed spin state change. If true, this is the first case when the YORP effect has been detected for a tumbling body.
\end{abstract}

\keywords{techniques: photometric -- minor planets, asteroids: individual (2012 TC4)}

\section{Introduction} \label{sec:intro}

Apollo-type near-Earth asteroid 2012~TC4 was discovered in October 2012 by the Pan-STARRS1 survey, few days before its closest approach to the Earth (having a geocentric distance of about 95,000\,km). It was observed photometrically and its rotation period of about 12\,min \citep{Pol:13, War:13b, Odd.ea:13, Car:14} and effective diameter of 7-34 m \citep{Pol:13} were determined. Later, \cite{Rya.Rya:17} noticed also a second period in the data and interpreted it as a manifestation of a tumbling rotation state.

The next close approach in October 2017 was at a geocentric distance of about 50,000\,km and an even more extensive observing campaign (including spectroscopic and radar observations) was coordinated by the NASA Planetary Defense Coordination Office (PDCO) at that time. This campaign served also as a planetary defense exercise and its results were summarised by \cite{Red.ea:19}. Additionally, \cite{Ura.ea:19} also conducted the observing campaign of this asteroid in the same apparition independently and they attempted to reproduce their light curves with a model of a tumbling triaxial ellipsoid. Besides these observing campaigns, a few photometric observations of 2012 TC4 were carried out in this apparition \citep{Son.ea:17, War:18, Lin.ea:19}. All photometric data observed in 2017 confirmed the excited rotation state of 2012~TC4 with the main period of 12.2\,min.

Here, we revisit the situation using more sophisticated methods and tools. We reconstruct the convex shape model and spin state of 2012 TC4 from the available light curves that include the published data in the literature and our own new observations. We show that the rotation state must have changed between 2012 and 2017 apparitions and we propose a YORP-driven spin evolution as the most likely explanation. The data are described in Sect.~\ref{sec:data}, the physical model of the body is developed in Sect.~\ref{sec:physical_model}, and the theoretical analysis of rotation dynamics is in Sect.~\ref{sec:theory}. Mathematical methods and numerical set-up of the theoretical model are summarized in the Appendix~\ref{methods}. The best fit of our physical model to the available light curves is shown in the Appendix~\ref{sec:lc_fits}.

\section{Optical Photometry Data}
\label{sec:data}

To reconstruct the spin state of 2012~TC4, we collected its light curves observed during both close approaches. Photometric observations from 2012 and 2017 were made using a variety of telescopes having apertures between 0.35~m and 5~m and equipped with CCD cameras. Observational circumstances with references to original sources are listed in Table~\ref{table:observation}.
Apart from previously published light curves, our dataset
also includes several new observations (indicated by coauthor names in the last column). 
In particular, we obtained four light curves using Pistoiese 0.6~m telescope (MPC code: 104) with CCD Apogee U6 which has a $35' \times 35'$ field of view corresponding to a pixel scale of 2 arcsec/pixel in both apparitions. The raw frames were processed for the dark and flat-field correction and the light curves of this observatory were constructed using Canopus software \citep{Warner2006}. 
The pre-processing and the photometry of the light curve observed at Wildberg Observatory (MPC code: 198) using a 0.35~m telescope equipped with SXVF-H16 $2048 \times 2048$ CCD Camera was conducted using Astrometrica \citep{Raab_2012}. In the pre-process for this data, both dark and flat-field correction was carried out. All photometric data were calibrated referenced the PPMXL Catalog \citep{Roeser_et_al_2010}.
The Panchromatic Robotic Optical Monitoring and Polarimetry Telescopes (PROMPT) located at the Cerro Tololo Inter-American Observatory (CTIO) in Chile consist of six 0.41-m reflectors equipped with the Apogee Alta U47+ E2V camera. The field of view is $10'\times 10'$ with 0.59 arcsec/pixel. All raw image frames were processed (master dark, master flat, bad pixel correction) using the software package MIRA. Aperture photometry was then performed on the asteroid and three comparison stars. A master image frame was created to identify any faint stars in the path of the asteroid. Data from images with background contamination stars in the asteroid's path were then eliminated.
A part of the published light curves, we obtained from the Asteroid Lightcurve Data Exchange Fromat database (ALCDEF\footnote{http://alcdef.org/}, \citealp{War.ea:11}). 
We also acquired the light curve published by \citep{Red.ea:19} from International Asteroid Warning Network (IAWN) 2012 TC4 Observing Campaign homepage\footnote{https://2012tc4.astro.umd.edu/Lightcurve/supplement/2012TC4\_Lightcurve\_Observations\_Summary.html}.

Since the corrected light curves were observed using various filters, and include both the relative and the absolutely calibrated observations, they have a magnitude offset between each other. As a result, the whole dataset is primarily treated as an ensemble of relative light curves. 

\subsection{Two-period Fourier series analysis}

In the first step, we analyzed the photometry data from 2012 and 2017 using the 2-period Fourier series method \citep{Pra.ea:05, Pra.ea:14}.
Concerning the 2012 observations, we used all but one photometric light curve series taken from 2012-10-09.9 to 2012-10-11.2 (see Table~\ref{table:observation}). In particular, we excluded the data taken on 2012-10-10.7 by \cite{Pol:13}, where we found a possible timing problem.\footnote{Indeed, David Polishook (personal communication) checked his data upon our request and confirmed that there was an issue with the times in his 2012 observations. We then used the corrected data for physical model reconstruction in Sect.~\ref{sec:physical_model}.}
Concerning the 2017 observations, we used a selected subset of the data taken between 2017-10-09.1 and 2017-10-11.1. This choice was motivated by noting that the observing
geometry during this time interval in 2017 was very similar (due to the fortuitous resonant return of the asteroid to Earth after the five years) to the geometry of the time interval of the 2012 observations. This choice minimizes possible (but anyway small) systematic effects due to changes in observing geometry when comparing results of our analysis of the data from the two apparitions (see below).

\begin{figure*}[t]
   \includegraphics[width=0.49\textwidth]{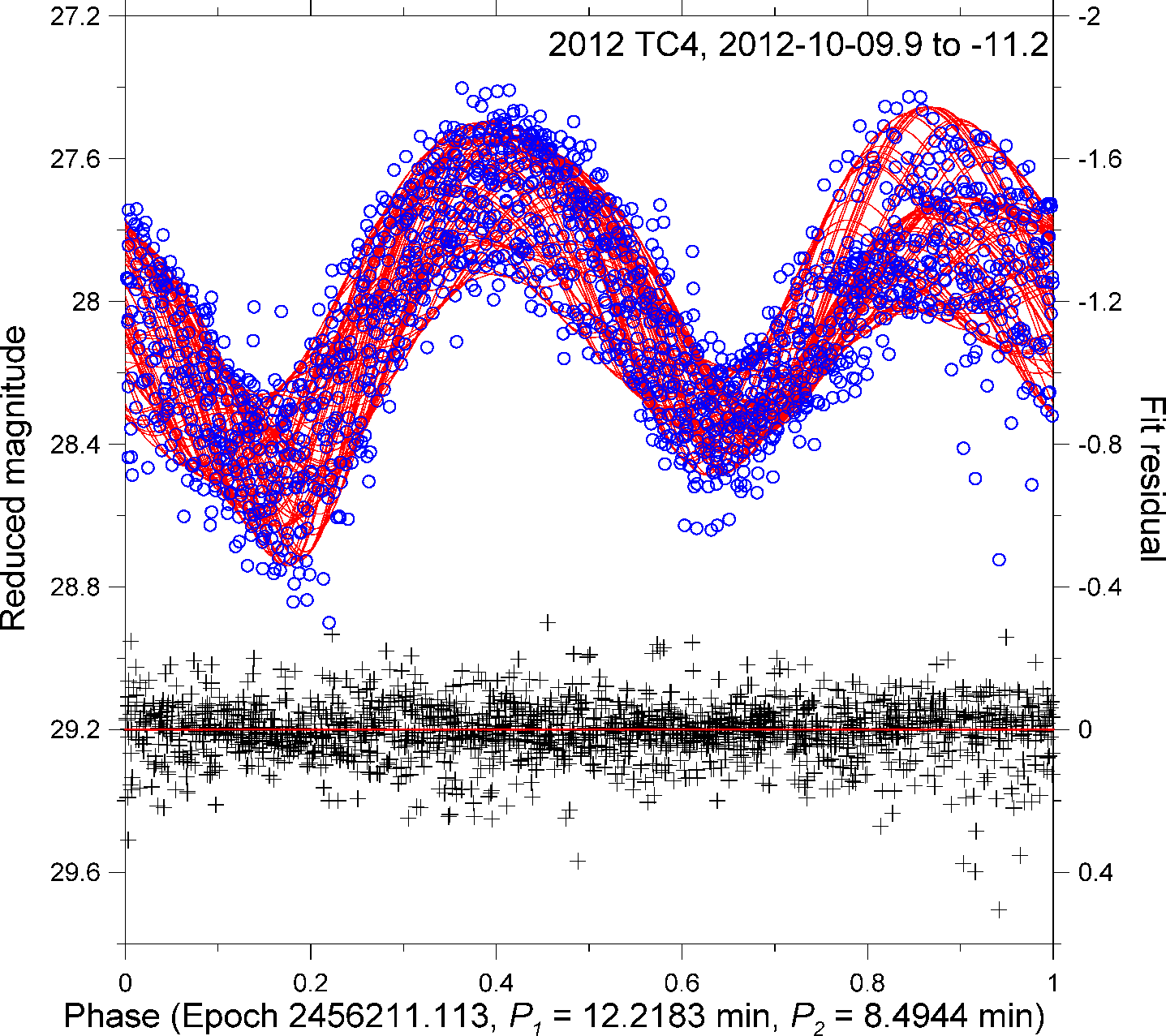}
   \includegraphics[width=0.49\textwidth]{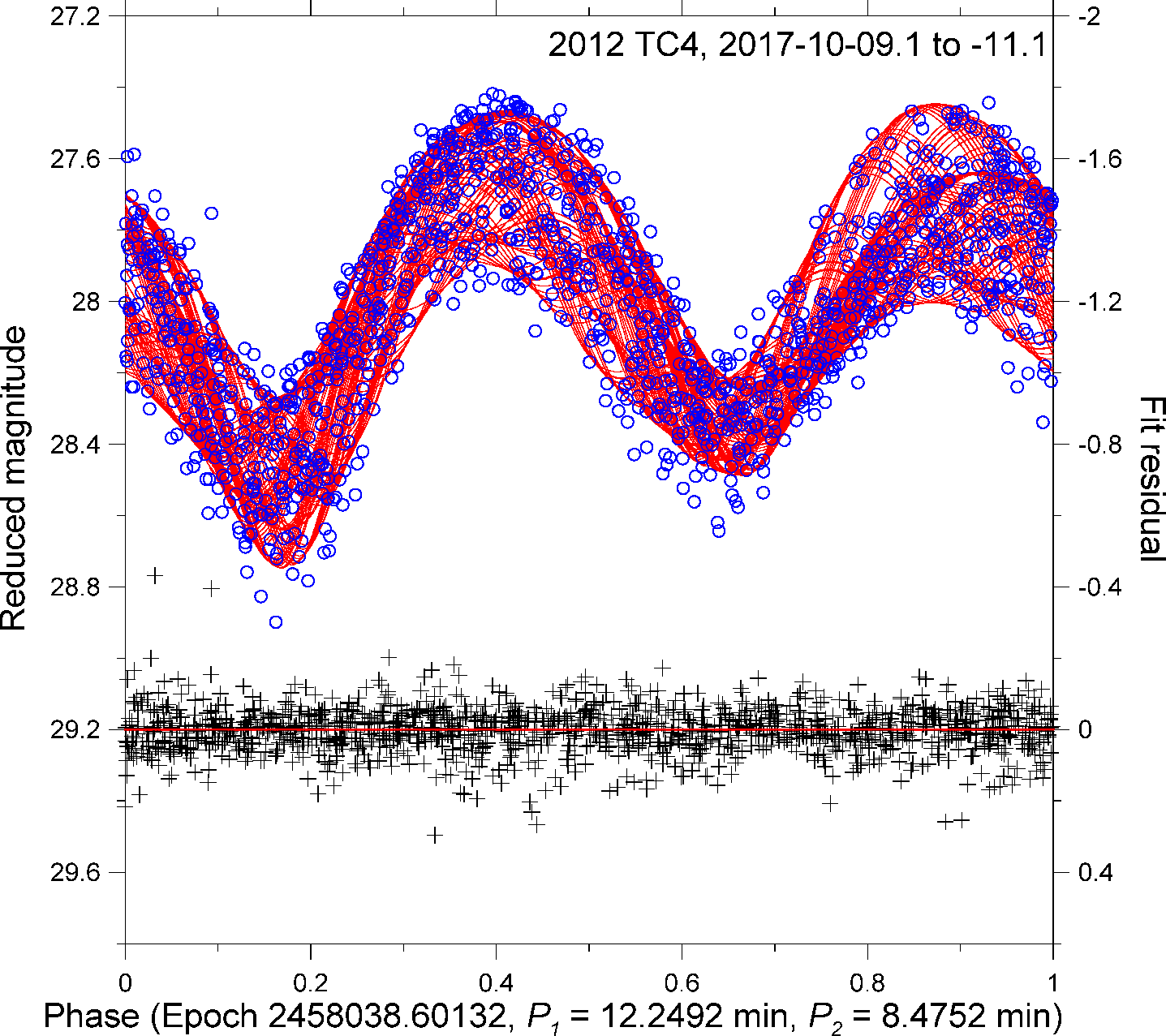}
   \caption{The blue open circles are the photometric data of 2012~TC4 taken from 2012-10-09.9 to 2012-10-11.2 (left) and from 2017-10-09.1 to 2017-10-11.1 (right) reduced to the unit geo- and heliocentric distances and to a consistent solar phase (see text for details), folded with the respective main periods $P_1$. The red curve is the best-fit 4th order Fourier series with the two periods. The black plusses are the post-fit residuals (see the right ordinates).}
   \label{fig:Fourier_fit}
  \end{figure*}

We reduced the data to the unit geo- and heliocentric distances and to a consistent solar phase using
the $H$-$G$ phase relation, assuming $G = 0.24$, converted them to flux units and fitted them
with the 4th order 2-period Fourier series. A search for periods quickly converged and we found the two main periods 
$P_1 = 12.2183 \pm 0.0002$\,min and $P_2 = 8.4944 \pm 0.0002$\,min in 2012 and
$P_1 = 12.2492 \pm 0.0001$\,min and $P_2 = 8.4752 \pm 0.0001$\,min in 2017.
The phased data and the best-fit Fourier series, together with the post-fit residuals, are plotted in Fig.~\ref{fig:Fourier_fit}.
We note that the smaller formal errors of the periods determined from the 2017 data were due to a higher quality of the 2017 observations 
(the best-fit rms residuals were 0.081 and 0.065~mag for
the 2012 and 2017 data, respectively; see also the post-fit residuals plotted at the bottom part of Fig.~\ref{fig:Fourier_fit}). As for possible systematic errors of the determined periods, the largest could be due to the so-called synodic effect. It is caused by the change of position of a studied asteroid with respect to the Earth and Sun in the inertial frame during the observational time interval. An estimate of the magnitude of the synodic effect can be obtained using the phase-angle-bisector approximation, for which we used eq.~(4) from \cite{Pra.ea:96}. 
Using this approach, we estimate the systematic errors of the determined periods    
$\Delta P_1 = 0.0005$ and 0.0002\,min and $\Delta P_2 = 0.0002$ and 0.0001\,min 
in 2012 and 2017, respectively. These systematic errors are only slightly larger than the formal errors given above. A caveat is that the formula eq.~(4) in \cite{Pra.ea:96} was determined for the case of a principal-axis rotator. An exact estimate of the systematic period uncertainties due to the synodic effect for a tumbling asteroid would require an analysis of its actual non-principal-axis rotation in given observing geometry, but we did not pursue it here as the effect was naturally surmounted by the physical modeling presented in the next section.    

As will be shown in the next section, the strongest observed frequency in the light curve, $P_1^{-1}$, is actually a difference between the precession and the rotation frequency of the tumbler: $P_1^{-1} = P_\phi^{-1} - P_\psi^{-1}$. The second strongest frequency $P_2^{-1}$ is then the precession frequency $P_\phi^{-1}$. This is a characteristic feature of light curve of a tumbling asteroid in short-axis mode (SAM; see below).
We note that the same behavior was observed for (99942) Apophis and (5247) Krylov that are also in SAM \citep{Pra.ea:14,Lee_et_al_2020}.

The principal light curve periods $P_1$ and $P_2$ determined from the 2012 and 2017 observations differ at a high level of significance, formally on an order of about $100\,\sigma$.
While the systematic errors due to the synodic effect could decrease the formal significance by a factor of a few, the significance
of the observed period changes would still remain large, on an order of several tens of $\sigma$. To interpret these findings in more depth, we turned to construct a physical model of 2012~TC4 as a tumbling object in the next section.                                                                             
    
\begin{table*}[t]
\caption{Observation details (new data denoted by name of a coauthor)}
\label{table:observation}
\begin{tabular}{l l l l}
\hline\hline
Telescope                                 & Date (UT)    & Filter         & Ref.     \\
\hline
\multicolumn{4}{c}{{\bf -- 2012 --}}                \\
OAVdA 0.81-m (Italy)                      & 2012 10 09.9 & C              & \cite{Car:14}     \\
Pistoiese 0.6-m (Italy)                   & 2012 10 09.9 & R              & Bacci   \\
Pistoiese 0.6-m (Italy)                   & 2012 10 10.0 & R              & Bacci   \\
MRO 2.4-m (USA)                           & 2012 10 10.2 & V              & \cite{Rya.Rya:17}     \\
Wise observatory 0.72-m (Israel)          & 2012 10 10.8 & V              & \cite{Pol:13}         \\
OAVdA 0.81-m (Italy)                      & 2012 10 10.8 & C              & \cite{Car:14}    \\
PROMPT1 0.41-m (Chile)                    & 2012 10 11.1 & Lum            & Pollock     \\
MRO 2.4-m     (USA)                       & 2012 10 11.1 & V              & \cite{Rya.Rya:17}     \\
PDO 0.35-m (USA)$^a$                      & 2012 10 11.2 & V              & \cite{War:13b}        \\ [4pt]
\multicolumn{4}{c}{{\bf -- 2017 --}}            \\
Kitt Peak Mayall 4-m (USA)                & 2017 09 13.2 & R              & \cite{Red.ea:19}      \\
Kitt Peak Mayall 4-m (USA)                & 2017 09 14.1 & R              & \cite{Red.ea:19}      \\
Palomar Hale 5-m (USA)                    & 2017 09 17.4 & SR             & \cite{Red.ea:19}      \\
Palomar Hale 5-m (USA)                    & 2017 09 20.2 & SR             & \cite{War:18}      \\
SOAR 4.1-m (Chile)                        & 2017 10 06.2 & SR             & \cite{Red.ea:19}      \\
PDO 0.35-m (USA)                          & 2017 10 09.2 & V              & \cite{War:18}      \\
MRO 2.4-m (USA)                           & 2017 10 09.2 & V              & \cite{Red.ea:19}      \\
Kiso 1.05-m (Japan)                       & 2017 10 09.5 & SG             & \cite{Ura.ea:19}      \\
Wise observatory 0.72-m (Israel)          & 2017 10 09.8 & V              & \cite{Red.ea:19}      \\
LCO-C 1-m (Chile)                         & 2017 10 10.1 & SR, SI         & \cite{Red.ea:19}      \\
LCO-A 1-m (Chile)                         & 2017 10 10.1 & SR             & \cite{Red.ea:19}      \\
PDO 0.35-m (USA)                          & 2017 10 10.2 & V              & \cite{War:18}         \\
Nayoro 0.4-m (Japan)                      & 2017 10 10.4 & V              & \cite{Ura.ea:19}      \\
BSGC 1-m (Japan)                          & 2017 10 10.6 & SG, SR, SI, SZ & \cite{Ura.ea:19}      \\
Lulin 1-m (Taiwan)$^b$                    & 2017 10 10.6 & BVRI (diff.)   & \cite{Lin.ea:19}      \\
Kiso 1.05-m (Japan)                       & 2017 10 10.5 & SG             & \cite{Ura.ea:19}      \\
Wise observatory 0.72-m (Israel)          & 2017 10 10.8 & V              & \cite{Red.ea:19}      \\
Pistoiese 0.6-m (Italy)                   & 2017 10 10.9 & R              & Bacci   \\
KMTNet 1.6-m (South Africa)               & 2017 10 10.9 & V              & \cite{Red.ea:19}      \\
Pistoiese 0.6-m (Italy)                   & 2017 10 11.0 & R              & Bacci   \\
USNA 0.51-m (USA)                         & 2017 10 11.0 & V              & \cite{Red.ea:19}      \\
MRO 2.4-m (USA)                           & 2017 10 11.1 & R              & \cite{Red.ea:19}      \\
PDO 0.35-m (USA)                          & 2017 10 11.2 & V              & \cite{War:18}         \\
Kiso 1.05 m (Japan)$^c$                   & 2017 10 11.5 & SG             & \cite{Ura.ea:19}      \\
Lulin 1-m (Taiwan)$^b$                    & 2017 10 11.6 & BVRI (diff.)   & \cite{Lin.ea:19}      \\
Anan Science Center 1.13-m (Japan)$^c$    & 2017 10 11.6 & V              & \cite{Ura.ea:19}      \\
Wise Observatory 0.72-m (Israel)          & 2017 10 11.8 & V              & \cite{Red.ea:19}      \\
AIRA 0.38-m (Romania)                     & 2017 10 11.8 & V              & \cite{Son.ea:17}      \\
Wildberg Observatory 0.35-m (Germany)$^d$ & 2017 10 11.8 &                & Apitzsch     \\
KMTNet 1.6-m (South Africa)               & 2017 10 11.9 & V              & \cite{Red.ea:19}      \\
MRO 2.4-m (USA)                           & 2017 10 12.1 & R              & \cite{Red.ea:19}      \\ [4pt]
\hline
\end{tabular}

\tablecomments{\scriptsize OAVdA: Astronomical Observatory of the Autonomous Region of the Aosta Valley, MRO: Magdalena Ridge Observatory, PROMPT1: Panchromatic Robotic Optical Monitoring and Polarimetry Telescopes, PDO: Palmer Divide Observatory, SOAR: Southern Astrophysical Research, LCO: Las Cumbres Observatory, BSGC: Bisei Spaceguard Center,  KMTNet: Korea Microlensing Telescope Network \citep{Kimetal2016}, USNA: The United States Naval Observatory, AIRA: Astronomical Institute of the Romanian Academy \\
$^a$ Split into six parts, each using different comparison stars. \\
$^b$ These data were estimated by subtracting the average magnitudes of the comparison stars. \\
$^c$ Split into two parts because we noted a possible calibration issue. \\
$^d$ Split into two parts, each using different comparison stars.}
\end{table*}

\section{Physical Model}
\label{sec:physical_model}

\subsection{Model from 2017 data}
\label{sec:model_2017}

The light curve data set from 2017 is much richer than that from 2012, so we started with the inversion of 2017 data. We investigated possible frequency combinations based on the fact that the main frequencies $f_1$ and $f_2$ of a tumbling asteroid light curve are usually found at 2$f_\phi$ and 2($f_\phi  \pm f_\psi$) or low harmonics and combination, where $f_\phi$ is the precession and $f_\psi$ the rotation frequency, respectively, and the plus sign is for a long-axis mode (LAM) and the minus sign for a short-axis mode (SAM) \citep{Kaa:01}. Using the values $f_1 = 4.898\,\text{hr}^{-1}$ and $f_2 = 7.079\,\text{hr}^{-1}$ from the previous section, we found 
eight possible frequency combinations: $f_1 = f_{\phi}$, $f_2 = 2 (f_{\phi} - f_{\psi})$ (SAM1); $f_1 = 2(f_{\phi} - f_{\psi}), f_2 = 2 f_{\phi}$ (SAM2); $f_1 = 2(f_{\phi} - f_{\psi}), f_2 = f_{\phi}$ (SAM3); $f_1 = f_{\phi} - f_{\psi}, f_2 = f_{\phi}$ (SAM4); $f_1 = 2 f_{\phi}, f_2 = 2 (f_{\phi} + f_{\psi})$ (LAM1); $f_1 = 2 f_{\phi}, f_2 = f_{\phi} + f_{\psi}$ (LAM2); $f_1 = f_{\phi}, f_2 = (f_{\phi} + f_{\psi})$ (LAM3); $f_1 = f_{\phi} + f_{\psi}, f_2 = f_{\phi}$ (LAM4). Then we conducted the shape and spin optimization for these combinations with the same way as in \citet{Lee_et_al_2020}. It was found that only the SAM4 solution provided an acceptable fit to the data and was physically self consistent.

We used 34 light curves from October 2017. Four light curves from September 2017 were very noisy and did not further constrain the model, so we did not include them in our analysis. We inverted the light curves with the method and code developed by \cite{Kaa:01} combined with Hapke's  light-scattering model \citep{Hapke_1993}. According to \cite{Red.ea:19}, colors of TC4 are consistent with C or X complex and also its spectrum is X type, with Xc type being the best match. The X complex contains low and high albedo objects but the E type seems most likely because the high circular polarization ratio suggests that TC4 is optically bright \citep{Red.ea:19}. This is in agreement with \cite{Ura.ea:19} who also report X type colors. As a result we used Hapke's model with parameters derived for an E type asteroid (2867)~\v{S}teins \citep{Spj.ea:12}: $\varpi = 0.57$, $g = -0.30$, $h = 0.062$, $B_{0} = 0.6$, and $\bar{\theta} = 28^\circ$. Because our data did not cover low solar phase angles, $h$ and $B_{0}$ parameters for opposition surge and also roughness $\bar{\theta}$ were fixed. We optimized $\varpi$ and $g$ parameters and they converged to values $\varpi = 0.69$, $g = -0.20$, which gave geometric albedo of 0.34. An alternative solution with fixed values $\varpi = 0.57$, $g = -0.30$ provided only marginally worse fit and the kinematic parameters were not affected. In general, solution of our inverse problem was not sensitive to Hapke's parameters likewise in previous studies \citep[e.g.,][]{Scheirich_et_al_2010, Pra.ea:14, Lee_et_al_2020}. 

The rotation and precession periods had the following values: $P_\psi = 27.5070 \pm 0.002$\,min and $P_\phi = 8.47512 \pm 0.0002$\,min, respectively. The $1 \sigma$ uncertainties were estimated from the increase of $\chi^2$ when varying the solved-for parameters -- given the number of measurements about 8600, $3\sigma$ uncertainty interval corresponds to about 5\% increase in $\chi^2$ \citep[e.g.,][]{Vok.ea:17}. Direction of the angular momentum vector in ecliptic coordinates was $\lambda = 92^\circ$, $\beta = -89.6^\circ$, practically oriented toward the south ecliptic pole. Normalized moments of inertia were $I_1 = 0.41$, $I_2 = 0.81$, but the model was not too sensitive to their particular values. The inertia moments computed from the 3D shape (assuming constant density) were $I_1 = 0.435$, $I_2 = 0.831$, indicating consistency with the kinematic parameters above. 

The dark facet, which is always introduced into a convex shape model to regularize the solution \citep{Kaa.Tor:01}, represented about a few percent of the total surface area and forcing it to smaller values led to a worse fit. This might mean that there is some albedo variegation on the surface of TC4 or that its real shape is highly nonconvex and a convex-shape approximation represent its limits.

\subsection{Model from 2012 data}
\label{sec:model_2012}

For this model, we used 14 light curves from 2012.  The rotation and precession periods were now $P_\psi = 27.873 \pm 0.005$\,min and $P_\phi = 8.4945 \pm 0.0003$\,min, respectively. The $1 \sigma$ uncertainties were estimated in the same way as for the 2017 model, namely from the increase of $\chi^2$. 
The direction of the angular momentum vector was $\lambda = 89^\circ$, $\beta = -90^\circ$, moments of inertia $I_1 = 0.43$, $I_2 = 0.80$. The 3D shape model was similar to that reconstructed from 2017 data and also the direction of the angular momentum vector was practically the same.

\begin{figure*}[t]
   \includegraphics[width=0.49\textwidth]{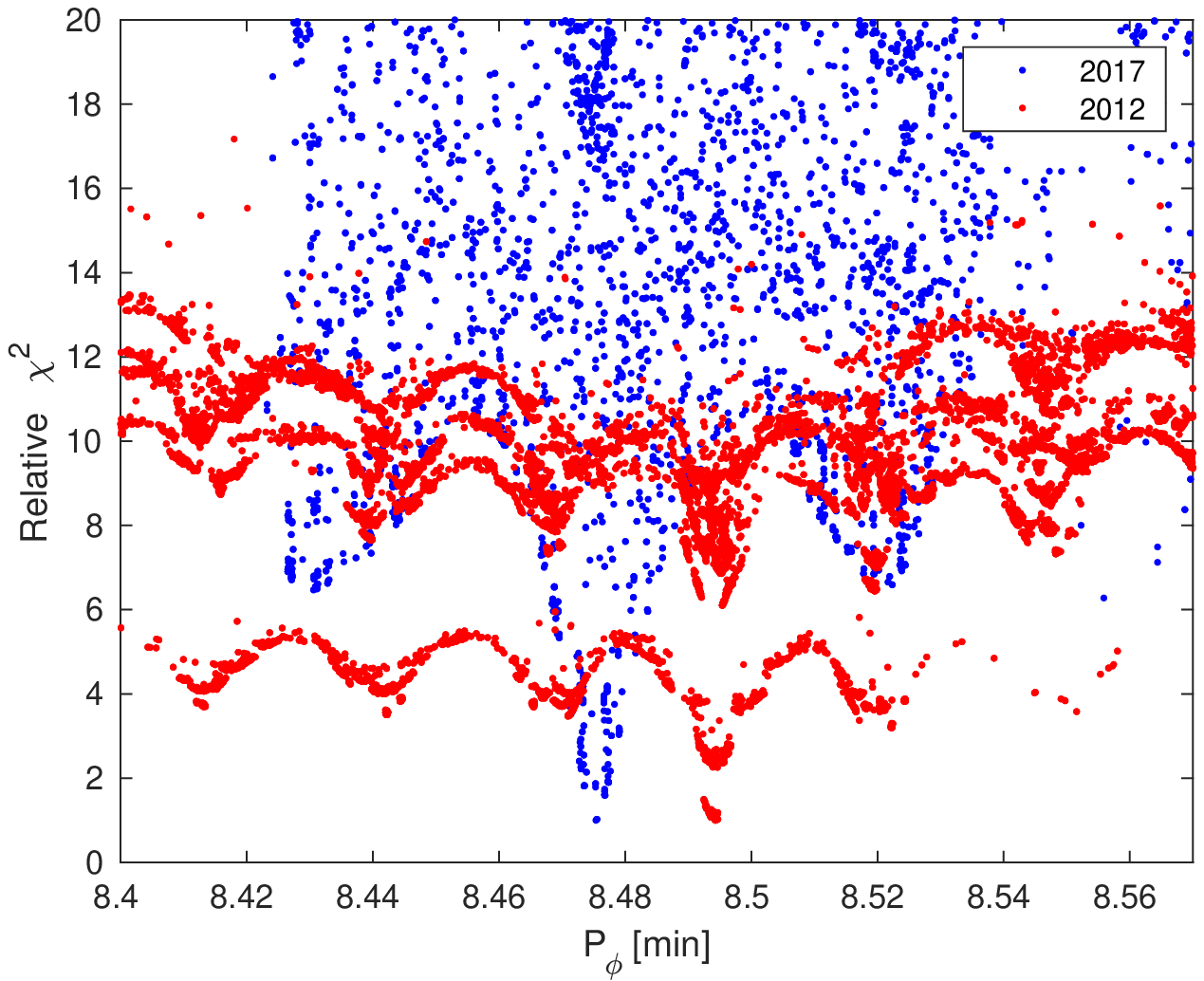}
   \includegraphics[width=0.49\textwidth]{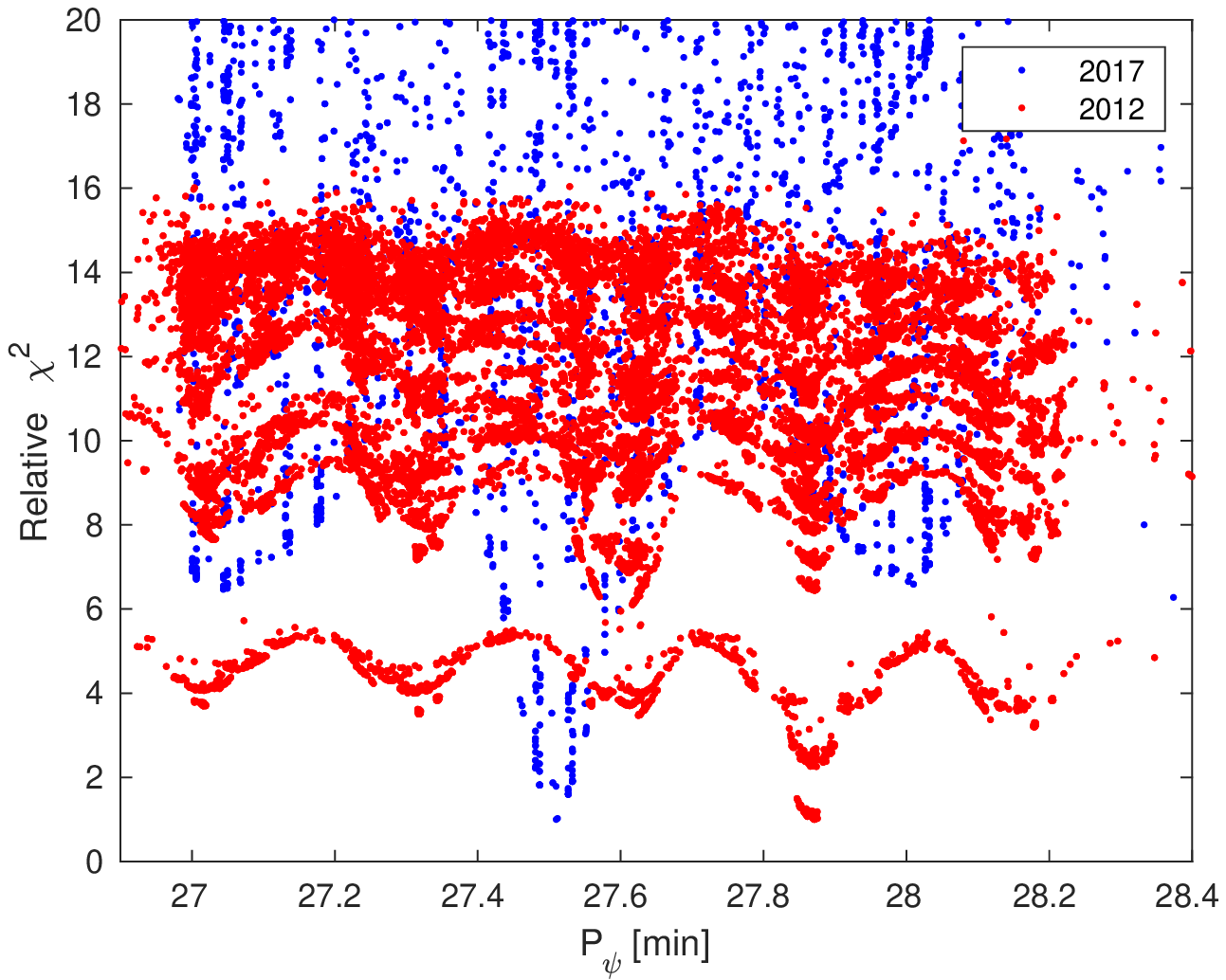}
   \caption{Period scan for data from 2017 (October only) and 2012. Each point represents one trial model that converged to given $P_\phi$, $P_\psi$, and $\chi^2$ values. Minima for 2012 and 2017 data are clearly separated. Relative $\chi^2$ was normalized to have the minimum value of 1.}
   \label{fig:period_scan}
  \end{figure*}

In spite of consistency in all other solved-for parameters, we thus observed that the periods $P_\phi$ and $P_\psi$ for the two apparitions were significantly different. Attempts to use the 2017 values to fit the 2012 data, or vice versa, led to a dramatically worse fit. We scanned period parameter space around the best-fit values and plotted the relative $\chi^2$ values -- see Fig.~\ref{fig:period_scan}. The minima in $\chi^2$ for 2012 and 2017 periods are clearly separated. 

\subsection{Model from 2012 and 2017 data}
\label{sec:model_2app}

  \begin{figure*}[t]
   \centering
   \includegraphics[width=0.7\textwidth, trim=2.5cm 8.5cm 1cm 1cm, clip]{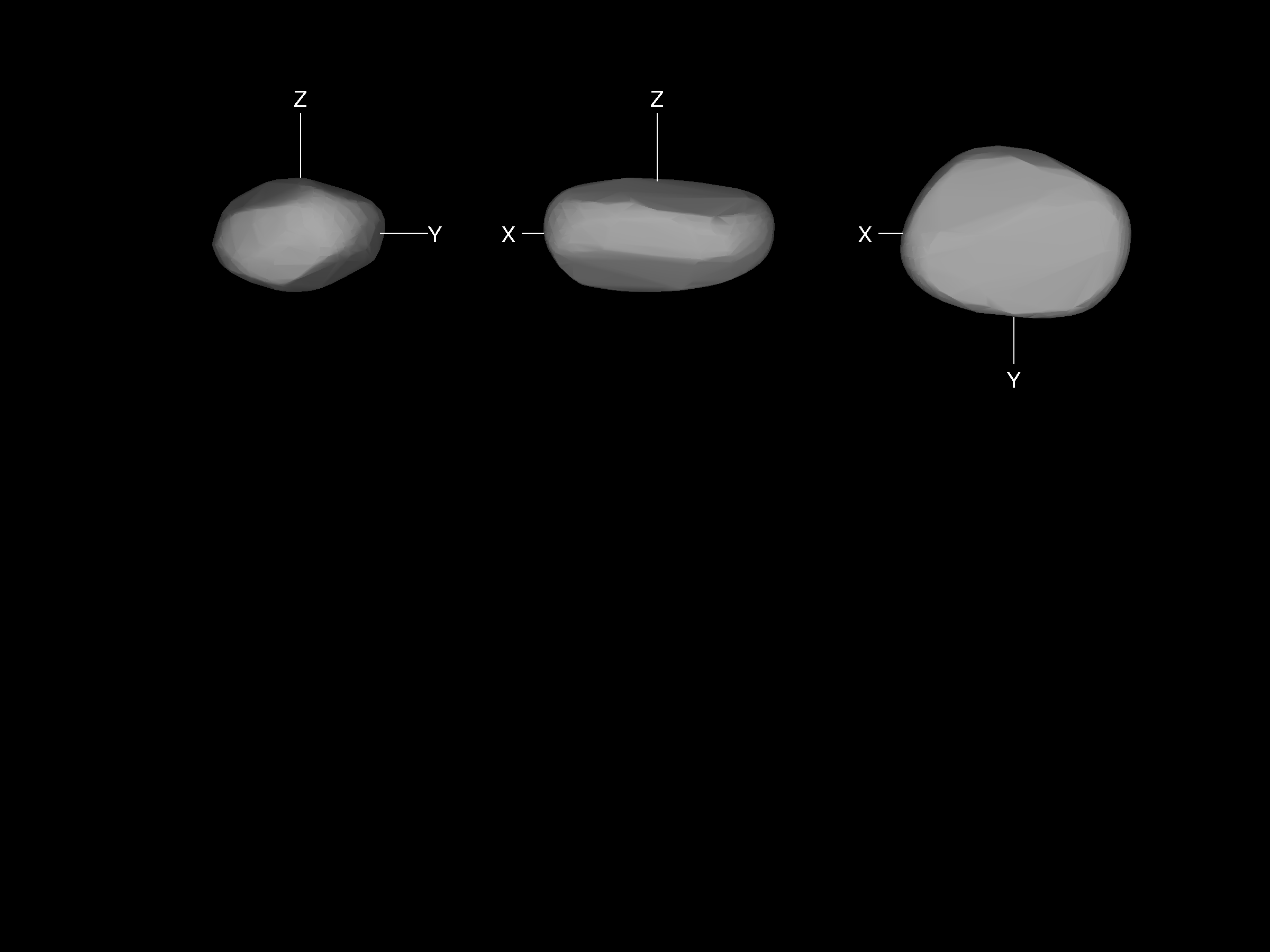}\\
   \includegraphics[width=0.7\textwidth, trim=2.5cm 8.5cm 1cm 1cm, clip]{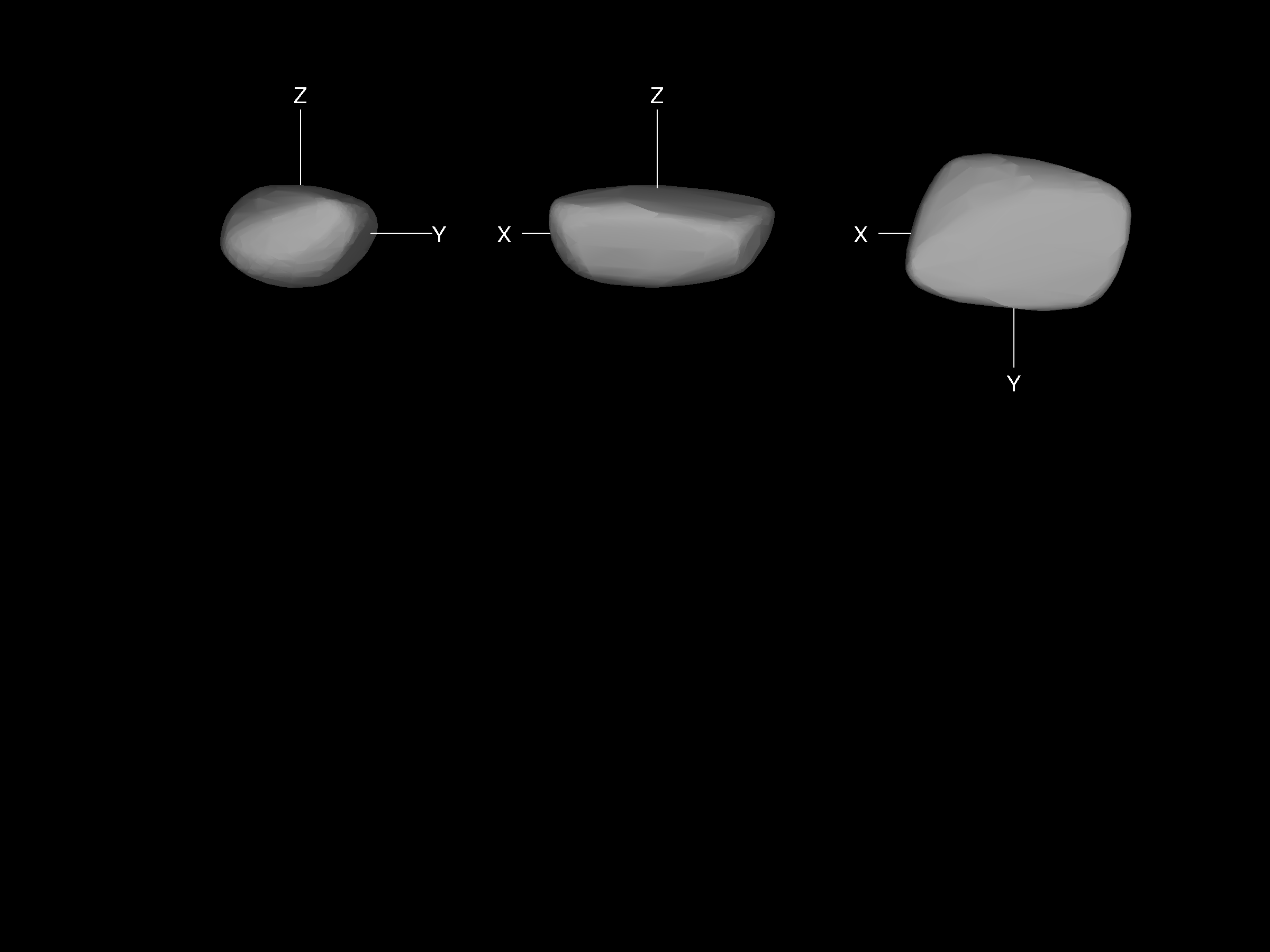}
   \caption{Shape models reconstructed independently from 2012 (top) and 2017 (bottom) light curves. The 2017 model is also almost identical to that reconstructed from joint inversion of 2012 and 2017 data.}
   \label{fig:shape_models_comparison}
  \end{figure*}

The 3D models reconstructed from the two apparitions independently have similar global shapes but their details are different (Fig.~\ref{fig:shape_models_comparison}). If we take the shape reconstructed from the 2017 data and use it to fit the 2012 data, it gives a satisfactory fit to light curves. However, a better way how to use all data together is not to treat them separately but to instead invert both apparitions together with a common shape model that would differ only in kinematic parameters. Therefore, we modified the original inversion code of \cite{Kaa:01} to enable including two independent light curve sets. We assumed that the only parameters that were different for the two apparitions were the rotation and precession periods $P_\psi$, $P_\phi$ and initial Euler angles $\phi_0$, $\psi_0$. Parameters describing the shape and the direction of the angular momentum vector were the same. So the full set of kinematic parameters for a two-apparition model was: $(\lambda, \beta, \phi_0^{(1)}, \phi_0^{(2)}, \psi_0^{(1)}, \psi_0^{(2)}, P_\psi^{(1)}, P_\psi^{(2)}, P_\phi^{(1)},  P_\phi^{(2)},I_1, I_2)$, where the superscript $(1)$ is for 2012 apparition and $(2)$ is for 2017 apparition. The two light curve data sets were independent in the sense that the integration of kinematic equations \citep[eq.~A.3 in][]{Kaa:01} was done separately for 2012 and 2017 data, the two epochs were not directly connected. The final model shape is almost identical to that reconstructed from only 2017 data shown in Fig.~\ref{fig:shape_models_comparison}. The fit of  the final model to individual light curves is shown in Figs.~\ref{fig:lc_fit_2012_1}--\ref{fig:lc_fit_2017_5}. Rotation and precession periods converged to practically the same values as with the independent treatment of each of two apparitions. The best-fit parameters for the 2012 and 2017 apparitions are listed in Table~\ref{tab:parameters}. The physical models of 2012 TC4 from 2012 and 2017 and the light curve data set used to reconstruct these models are available from the DAMIT database\footnote{https://astro.troja.mff.cuni.cz/projects/damit/} \citep{Dur.ea:10}.


\begin{table*}[t]
\caption{Parameters of the model in Fig.~\ref{fig:shape_models_comparison} reconstructed from 2012 and 2017 light curves. The reported errors correspond to standard deviations of parameters computed from bootstrap models. Parameters without errors were fixed (Hapke's parameters) or did not change ($I_1$ and $I_2$). Formally small uncertainty of $\psi_0$ means that this initial orientation angle is correlated with the shape and does not change significantly with different bootstrap data sets. Parameter $\lambda$ is practically unconstrained because the angular momentum direction is very close to $\beta = -90^\circ$. Values $\delta L/L$ and $\delta E/E$ are relative changes of angular momentum $L$ and energy $E$ between 2012 and 2017, i.e., $\delta L/L = (L_{2017} - L_{2012}) / L_{2017}$ and $\delta E/E = (E_{2017} - E_{2012}) / E_{2017}$.}
\label{tab:parameters}
\centering
\begin{tabular}{l c c}
\hline\hline
                & 2012                    & 2017 \\
\hline
$P_\psi$ [min]  & $27.8720 \pm 0.0007$    & $27.5070  \pm 0.0002\ $     \\
$P_\phi$ [min]  & $\ 8.4944 \pm 0.0005$   & $\ 8.47511 \pm 0.00008$    \\
$\phi_0$ [deg]  &   $322 \pm 8$           & $74 \pm 9$                 \\
$\psi_0$ [deg]  & $198 \pm 0.02$           & $180 \pm 0.2$          \\
JD$_0$          & 2456210.00    & 2458032.69    \\
\hline
$\lambda$ [deg] & \multicolumn{2}{c}{$103 \pm 78$}       \\
$\beta$ [deg]   & \multicolumn{2}{c}{$-88.5 \pm 0.7$}     \\
$I_1$           & \multicolumn{2}{c}{0.42}      \\
$I_2$           & \multicolumn{2}{c}{0.81}      \\
$w$             & \multicolumn{2}{c}{0.67}      \\
$g$             & \multicolumn{2}{c}{$-0.20$}      \\
$h$             & \multicolumn{2}{c}{0.062}      \\
$B_0$           & \multicolumn{2}{c}{0.6}      \\
$\bar{\theta}$ [deg] & \multicolumn{2}{c}{28}      \\
$\delta L/L$      & \multicolumn{2}{c}{$0.00078 \pm  0.00006$} \\
$\delta E/E$      & \multicolumn{2}{c}{$-0.0035 \pm 0.0001$}  \\
\hline
\end{tabular}
\end{table*}

 \subsection{Bootstrap} \label{boo}
 
\begin{figure*}[t]
   \includegraphics[width=0.49\textwidth]{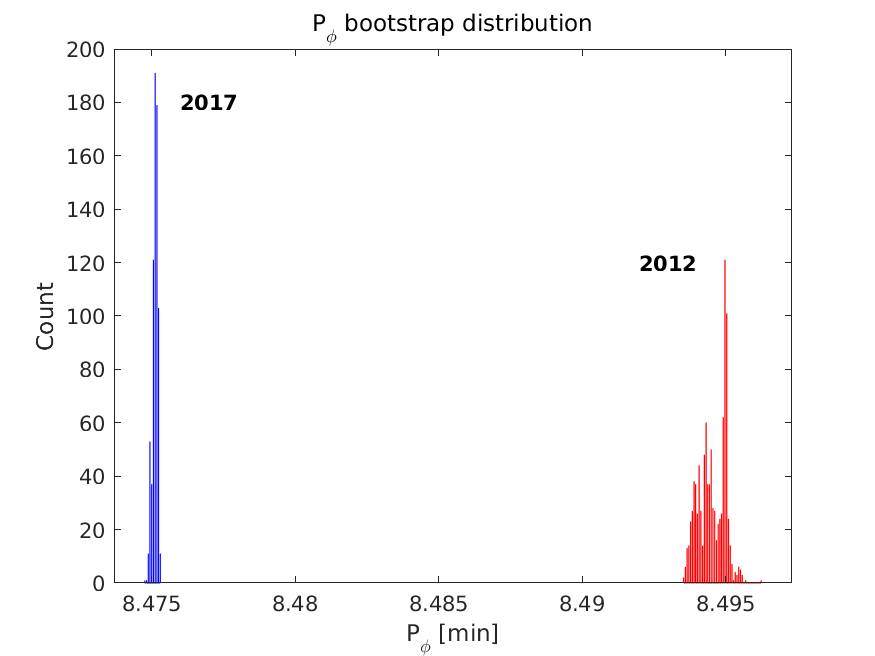}
   \includegraphics[width=0.49\textwidth]{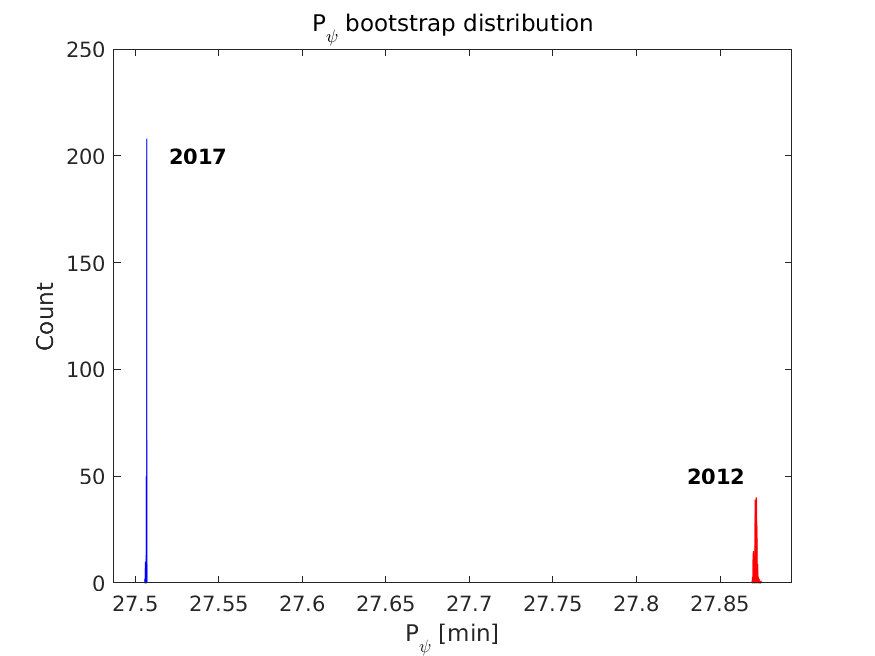} 
   \caption{Distribution of periods $P_\phi$ (left) and $P_\psi$ (right) for bootstrapped light curves.}
   \label{fig:histogram_periods}
  \end{figure*}

 To estimate uncertainties of physical periods, and to further robustly demonstrate that their change between 2012 and 2017 is significant, we created bootstrapped data samples and repeated the light curve inversion for them. From both 2012 and 2017 data sets, we created 1000 bootstrap samples by randomly selecting the same number of light curves from the original data set. For October 2017 bootstrap, 279 final shape models had clearly wrong inertia tensor that was not consistent with the kinematic $I_1, I_2$ parameters, so we removed them from the analysis. For all remaining bootstrap models, we plot histograms of $P_\phi$ and $P_\psi$ distribution in Fig.~\ref{fig:histogram_periods}. The standard deviations of the precession period $P_\phi$ are $0.0005\,$min and $0.00009\,$min for 2012 and 2017 data, respectively, which are similar to uncertainty values derived in Sects.~\ref{sec:model_2017} and \ref{sec:model_2012}. For the rotation period $P_\psi$, these standard deviations are $0.0008$ and $0.0002$\,min, which is significantly smaller than our $\chi^2$-based estimate. Nevertheless, the difference between periods determined from 2012 and 2017 apparitions is much larger than their uncertainty intervals and we are not aware of any random or model errors that could cause such difference. Our conclusion is that the spin state of TC4 has changed from 2012 to 2017, rotation and precession periods have decreased. In what follows, we try to interpret this change using a theoretical model of TC4 spin evolution with the relevant torques.
 
\section{Theory}
\label{sec:theory}

\subsection{Orbital dynamics}

\begin{figure}[tp]
\begin{center}
 \includegraphics[width=12cm]{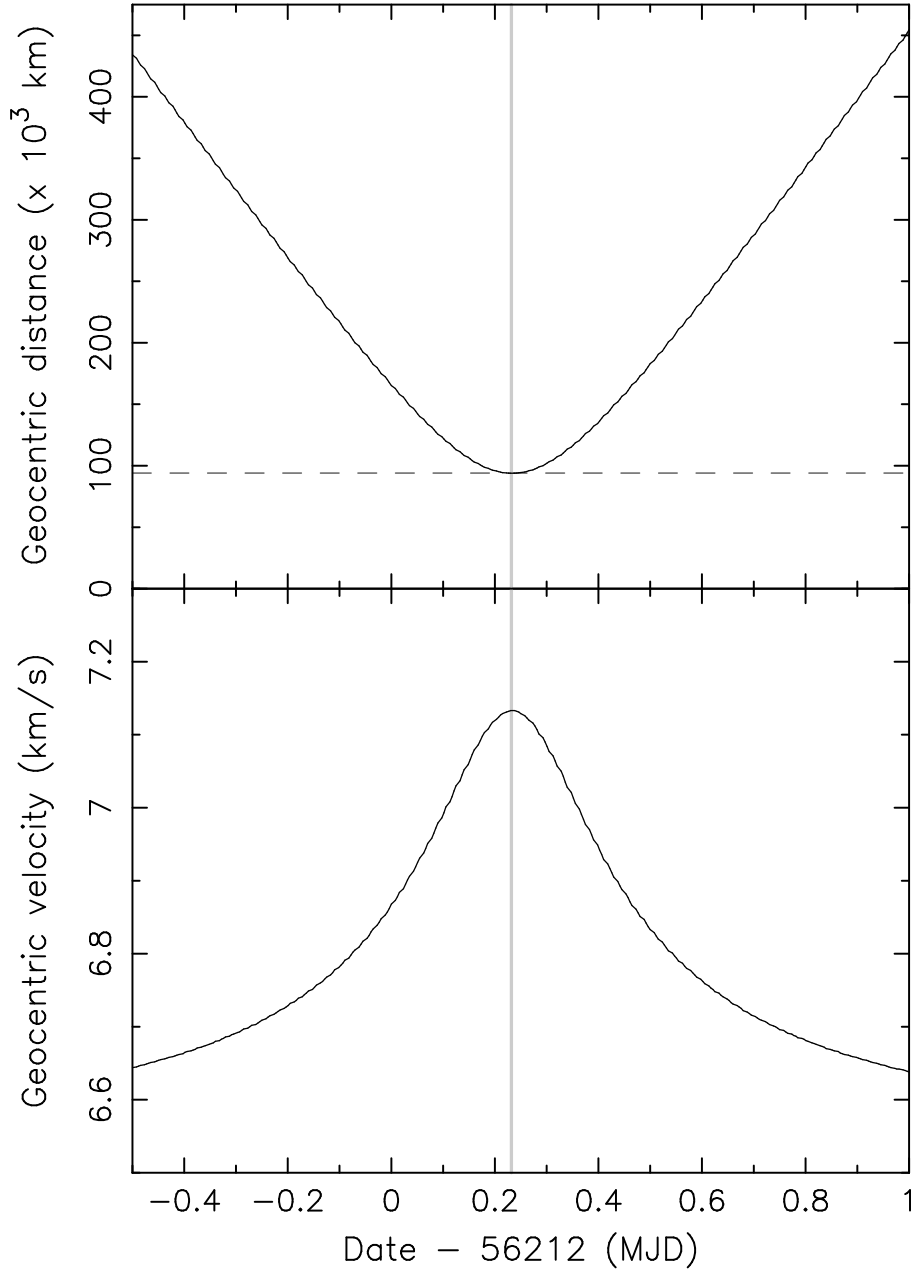}
\end{center}
\caption{Geocentric distance (top) and relative velocity (bottom)
 of 2012~TC4 during its close encounter with Earth on October~12,
 2012 (nominal minimum configuration at MJD56212.229, gray vertical
 line); the abscissa shows time in days with respect to MJD56212.
 The dashed horizontal line at the top panel shows the minimum distance
 of $\simeq 95,000$~km. The asymptotic value of the relative velocity
 with respect to the Earth, $\simeq 6.5$~km/s, increases to more than
 $7.1$~km/s by the Earth gravity.}
\label{f1}
\end{figure}

The unique observational opportunities of 2012~TC4 are directly related to its
exceptional orbit. The asteroid had a deep encounter with the Earth on October~12, 2012, 
during which the closest distance to the geocenter was approximately $95,000$~km
(Fig.~\ref{f1}). 
However, the more unusual circumstance was that the 2012 close encounter resulted
in a change of the 2012~TC4's orbital semimajor axis which placed it nearly
exactly to the 5:3 resonance with the Earth heliocentric motion. As a result,
in five years after the first close encounter, i.e. on October~12, 2017, the relative 
configuration of the asteroid and the Earth nearly exactly repeated, placing it again
in a deep encounter configuration. This time the closest approach to the geocenter
had even closer distance of $50,200$~km. Astrometric observations during 
the two close approaches, including Arecibo and Green Bank radar data taken in 2017,
allowed a very
accurate orbital solution over the five year period of time in between 2012
and 2017. In the context of this paper, we note that it also provided an interesting 
information about nongravitational effects needed to be empirically included in orbit
determination. Adopting methodology from cometary motion (e.g., \cite{Mars.ea:73} and 
\cite{Far.ea:13} or \cite{Mom.ea:14} for the asteroidal context),
we note the following values of radial $A_1$ and transverse $A_2$ accelerations: 
(i) $A_1 = (2.17\pm 0.80)\times 10^{-11}$ au~d$^{-2}$, or $(4.35\pm
1.60) \times 10^{-10}$ m~s$^{-2}$, and (ii) $A_2 = -(2.73\pm 0.65)\times
10^{-13}$ au~d$^{-2}$ (both assume $\propto r^{-2}$ heliocentric decrease;
see JPL/Horizons web page \url{https://ssd.jpl.nasa.gov/sbdb.cgi}). At a first sight,
these values appear very reasonable \citep[compare, e.g., with a $A_1$ and
$A_2$ fits for a $\simeq 4$~m body 2009~BD,][]{Mom.ea:14,Vok.ea:15}.

If we were to interpret both components as a result of radiation
forces, we may further obtain useful information about the body. The
radial component would represent the direct solar radiation pressure.
In a simple model, where we assume a spherical body of size $D$ and bulk density
$\rho$, we have $A_1\simeq  3C_{\rm R} F_0/(2\rho D c)$, with $C_{\rm R}$
the radiation pressure coefficient (dependent on sunlight scattering
properties on the surface), $F_0\simeq 1367$ W~m$^{-2}$ the solar constant and
$c$ the light velocity. Adopting $C_{\rm R}\simeq 1.2$ and $D\simeq 10$~m,
we obtain a very reasonable bulk density $\rho\simeq (1.9\pm 0.7)$ g~cm$^{-3}$.
The transverse component of nongravitational orbital effects makes sense
when interpreted as the Yarkovsky effect \citep[e.g.,][]{Vok.ea:15}.
Note that the above mentioned value of $A_2$ translates to a secular 
change of the semimajor axis $da/dt = -(110\pm 26)\times 10^{-4}$ au~My$^{-1}$ \citep[see, 
e.g.,][]{Far.ea:13}. Given the near extreme obliquity of the 
rotational angular momentum, we may safely restrict to the diurnal component 
of the Yarkovsky effect. The negative value of $A_2$, or $da/dt$, corresponds 
well to the retrograde sense of 2012~TC4 rotation (implied by the direction of
the rotational angular momentum vector, Sec.~\ref{sec:physical_model}). Next, borrowing the simple model for
a spherical body from \cite{Vok:98}, and fixing the size $D=10$~m, the
inferred bulk density $\rho\simeq 1.9$ g~cm$^{-3}$ and the surface thermal
conductivity $K=0.05$ W~m$^{-1}$~K$^{-1}$, we estimate that the 
corresponding surface thermal inertia is $\Gamma\simeq 490^{+270}_{-250}$ in SI units \citep[though, we note there is also a lower-inertia solution possible like in][]{Mom.ea:14}. This is a very adequate value too \citep[e.g.,][]{Del.ea:15}.
Obviously, due to many frozen parameters in the
model (and its simplicity), the realistic uncertainty in $\Gamma$ would be
larger. However, it is not our intention to fully solve this problem. We
satisfy ourselves with observation that the needed empirical nongravitational
accelerations in the orbital fit may be very satisfactorily interpreted as
radiation-related effects. In the next sections, we show that also the change
in the rotation state in between 2012 and 2017 observation epochs may be
very well explained by the radiation torque known as the 
Yarkovsky-O'Keefe-Radzievskii-Paddack effect \citep[e.g.,][]{Vok.ea:15}.

\subsection{Rotational dynamics}
\begin{figure}[tp]
\begin{center}
 \includegraphics[width=12cm]{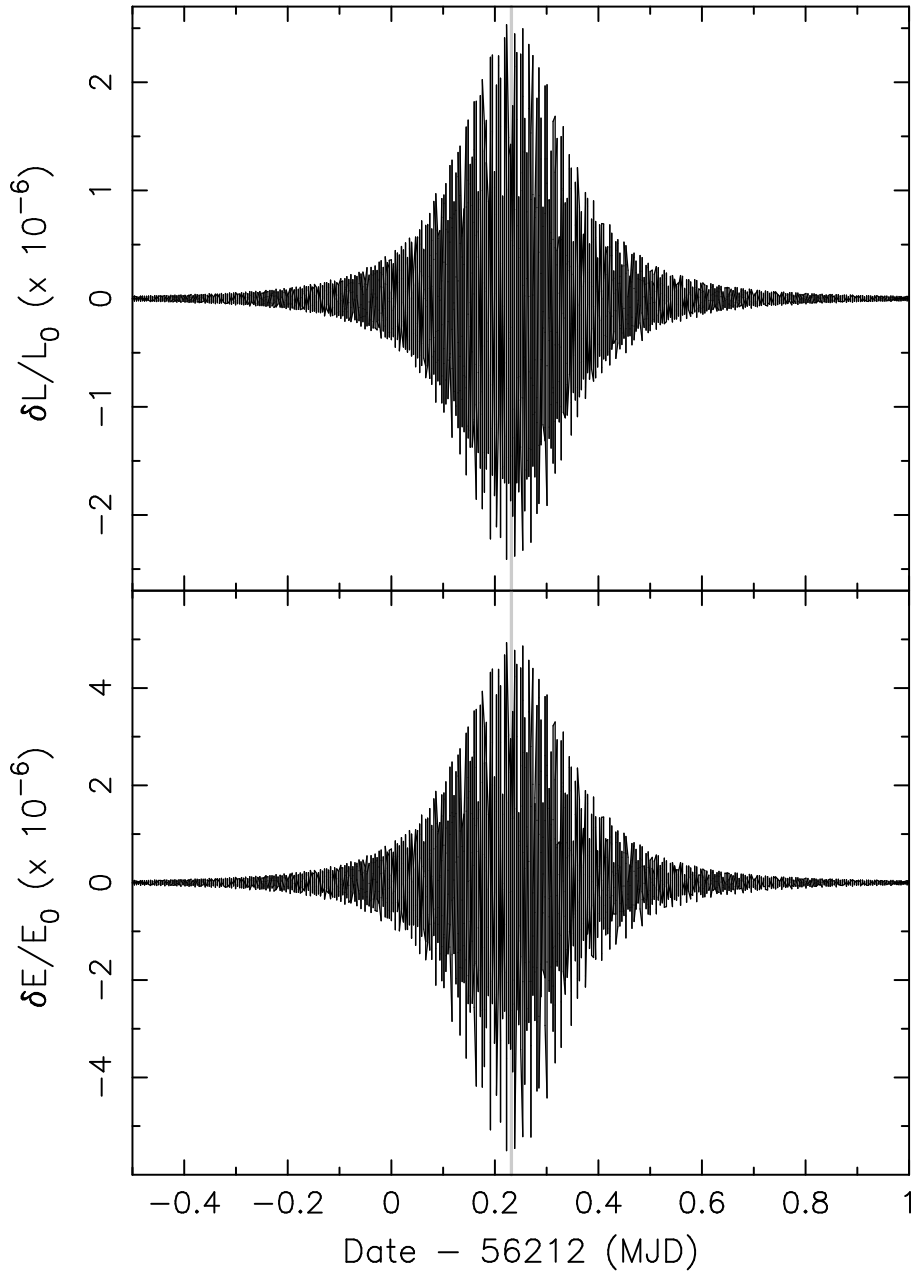}
\end{center}
\caption{The effect of the Earth gravitational torque in quadrupole
 approximation on rotation state parameters of 2012~TC4 during its close
 encounter on October~12, 2012 (gray line denotes the nominal minimum
 distance configuration). The abscissa shows time in days with respect
 to MJD56212 (as in Fig.~\ref{f1}). The upper panel shows a fractional
 change of the rotational angular momentum $\delta L=L-L_0$, normalized by 
 the initial value $L_0$, the bottom panel panel shows a fractional change
 of the rotational energy $\delta E = E - E_0$, normalized by the initial
 value $E_0$ (note that
 both ordinate scales are in $10^{-6}$). The resulting change of both
 parameters after the encounter is $\leq 5\times 10^{-9}$.}
\label{f2}
\end{figure}

Our methods and mathematical approach used to describe evolution of the rotation
state of 2012~TC4 are presented in the Appendix~\ref{methods}, therefore here we provide just a general outline. We numerically integrated Euler equations (\ref{e2}) and 
(\ref{e4}) describing spin state evolution of 2012~TC4 in between the 
observation runs in 2012 and
2017. The kinematical part, describing transformation between the
inertial frame and the body-frame defined by principal axes of the
tensor of inertia, was parametrized by the Rodrigues-Hamilton
parameters $\mbox{\boldmath$\lambda$}=(\lambda_0,\lambda_1,\lambda_2,\lambda_3)$
\citep[see, e.g.,][]{Whi:1917}. This choice helps to remove problems
related to coordinate singularity given by zero value of the nutation
angle. The dynamical part is represented by evolution of the angular velocity
$\mbox{\boldmath$\omega$}$ in the body-frame. Note that in most cases of
asteroid lightcurve interpretation, a simplified model of a free-top would
be sufficient (used also in Sec.~\ref{sec:physical_model} to fit the 2012 and 2017 data separately). However, the evidence of change of the rotation state of
2012~TC4 in between the 2012 and 2017 epochs, discussed above, requires appropriate
torques to be included in the model. We addressed two effects:
\begin{itemize}
\item gravitational torques due to the Sun and the Earth, and
\item radiation torques due to the sunlight scattered by the surface
 and thermally re-radiated \citep[the YORP effect; e.g.,][]{Bot.ea:06,Vok.ea:15}.
\end{itemize}
The tidal gravitational fields of the Sun and the Earth were represented
in the body-frame using the quadrupole approximation (\ref{e7}) \citep[e.g.,][]{Fit:70,Tak.ea:13}. The nature of the perturbation is different
for the Sun and the Earth. In the solar case, the gravitational torque results in
a small tilt by less than $1$ arcminute describing a small segment on the
precession cone. The effect of the Earth-induced gravitational torques
manifests as an impulsive effect only during close encounters. Our main
goal was to verify that, due to fast rotation of 2012~TC4, the effect
averages out and cannot contribute to the observed change in rotational
frequencies. Indeed, Fig.~\ref{f2} shows change of the osculating rotational
(intrinsic) angular momentum and energy during the 2012 close encounter
with the Earth (only about six times larger effects are observed during the closer
encounter in 2017, but this is not relevant for our analysis anyway, because
observations preceded this approach). The initial data of the simulation were taken
from the observations fit in 2012, namely before the encounter. Recall that
the characteristic timescale
of the encounter is about half a day (determined, for instance, as a width
of characteristic Earth relative velocity increase at the bottom panel of
Fig.~\ref{f1} at $\simeq 6.8$~km s$^{-1}$ level, half value between the asymptotic and
peak velocities). In comparison, the characteristic rotational periods $P_\psi$
and $P_\phi$ are of the order of minutes, i.e. much shorter. As a result,
the effect of the Earth gravitational torque efficiently averages out during
each of the rotation cycles \citep[a significant effect may be expected only for
very slowly rotating bodies and sufficiently deep encounters, such as
seen for 4179 Toutatis during its 2004 close encounter with the Earth; e.g.,][]{Tak.ea:13}. Therefore, we may conclude that the gravitational
torques cannot explain the observed change in the rotation state of 2012~TC4
in between the 2012 and 2017 epochs. Still, we keep them in our model for
the sake of completeness.
\begin{figure}[tp]
\begin{center}
 \includegraphics[width=12cm]{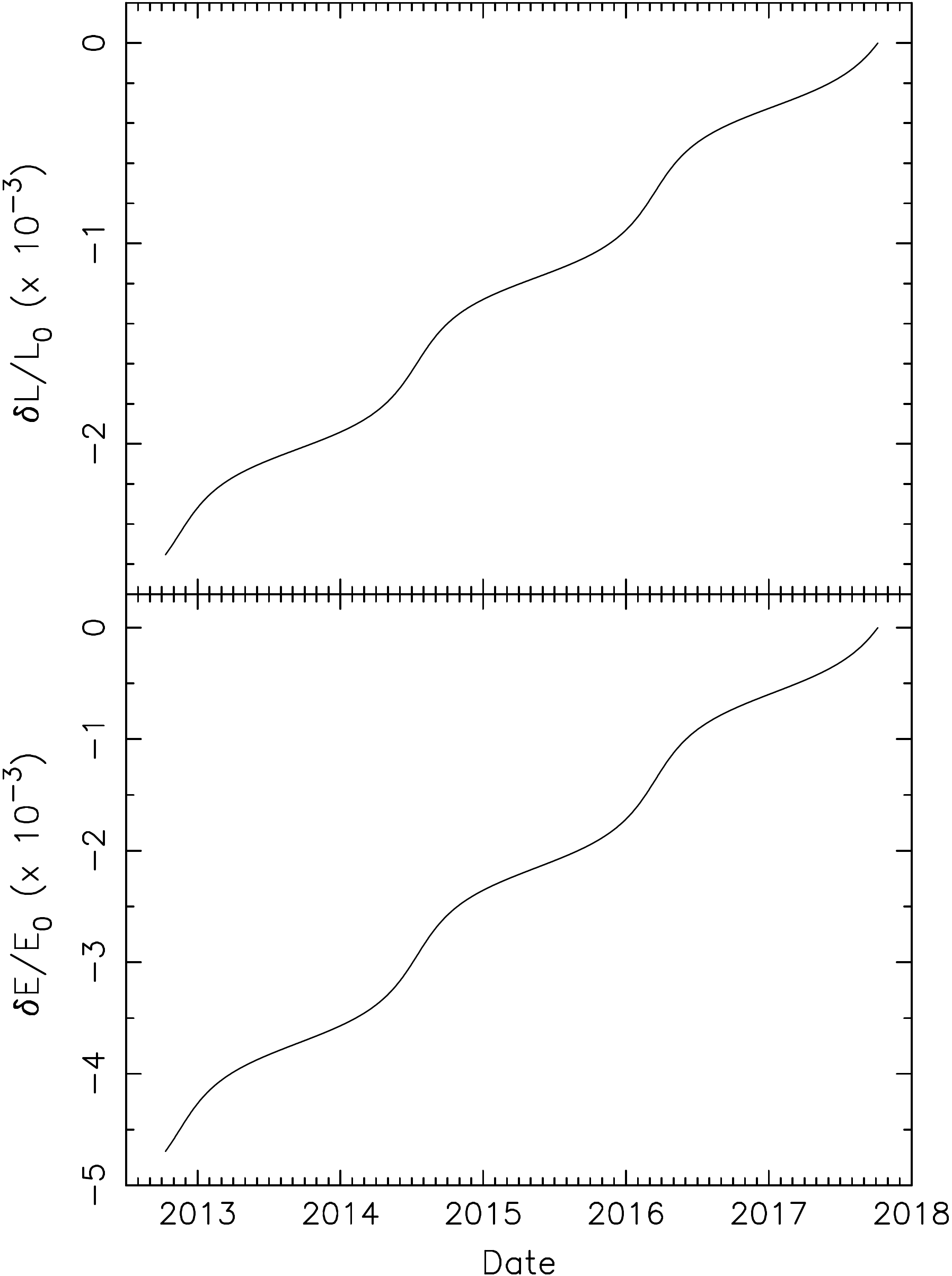}
\end{center}
\caption{The effect of the radiation (YORP) torque on rotation state parameters
 of 2012~TC4 in the time interval between the two recent close encounters with
 the Earth (i.e., October~12, 2012 and October~12, 2017). Nominal model is used here to
 obtain a first insight of the expected order of magnitude of the perturbation.
 The abscissa shows time in years. The upper panel shows a fractional
 change of the rotational angular momentum $\delta L=L-L_0$, normalized by 
 the initial value $L_0$, the bottom panel panel shows a fractional change
 of the rotational energy $\delta E = E - E_0$, normalized by the initial
 value $E_0$ (note that both ordinate scales are in $10^{-3}$). Both reference
 values taken at MJD58032.19, mean epoch of 2017 observations, and the asteroid's
 rotation state was propagated backward in time to October 2012.}
\label{f3}
\end{figure}

The radiation torques are of a quite different importance. It is well-known
that the YORP effect is able to secularly change rotational frequency and
tilt the rotational angular momentum in space. While mostly studied in the
limit of a rotation about the shortest axis of the inertial tensor, generalizations
to the tumbling situation were also developed. Both numerical \citep{Vok.ea:07}
and analytical \citep{Cic.Sch:10, Bre.ea:11} studies
confirmed that YORP effect is able to change rotational angular momentum, its
orientation in both the inertial space and body-frame in an appreciable manner.
As usual, the effect is more important on small bodies. Here we included the
simple, zero-inertia limit developed in \cite{Rub:00} and \cite{Vok.Cap:02}, 
see Eq.~(\ref{e8}). A first look into importance of the radiation
torques may be obtained by taking nominal solution of the rotation state from
the 2017 observations, including the appropriate shape model, and propagate it 
backwards in time to early October 2012 (epoch of the first set of observations). 
We use the more accurate 2017 model as reference rather than the one constructed 
from poorer 2012 observations. The length-scale of the model was adjusted to
correspond to an equivalent sphere of diameter $D=10$~m and density was
$\rho=1.4$ g~cm$^{-3}$. We verified that results have the expected invariance to
rescaling of both $\rho$ and $D$ such that $\rho\,D^2=$~const. Our nominal choice of the $1.4$ g~cm$^{-3}$ bulk density, albeit conflicting with the suggested E-type specral classification of TC4, is therefore linked to the assumed equivalent size of $10$~m, but it might
be redefined according to the rescaling principle. This combination of parameters provides a very nice match of the $P_\phi$ period change due to our YORP model (see Sec.~\ref{sec:theory_res}).

Figure~\ref{f3} provides
information about secular change in rotational angular momentum $L$ (top) and
energy $E$ (bottom) in that simulation. Here we see a long-term change in both
quantities. The wavy pattern is due to eccentricity of the 2012~TC4 orbit
and a stronger YORP torque at perihelion. The accumulated fractional change in both $L$
and $E$ is few times $10^{-3}$. This is promising, because the observed change in these quantities in of the same order of magnitude (see Table~\ref{tab:parameters}), and makes us believe that the change in the directly observable $P_\psi$ and $P_\phi$ periods will also be as needed.
Nevertheless, we also note a difference. The simulation results shown in Fig.~\ref{f3} indicate both angular momentum and energy increased from 2012 to 2017. Such a behaviour is perhaps expected at the first place (for instance, in the case of a body in a principal-rotation state, YORP would necessarily affect both $E$ and $L$ in the same way). However, rotation state solutions from observations in 2012 and 2017 tell us something else (see Table~\ref{tab:parameters}): the rotational angular momentum $L$ increased in between 2012 and 2017, while the energy $E$ decreased. Before commenting more on this difference, we first provide more detailed analysis of the radiation torque effects for 2012~TC4 in our modeling, this time using the whole suite of acceptable initial data and shape models (all compatible with the observations; Sec.~\ref{boo}). This will allow a statistical assessment of the predicted values.

\subsection{Results} \label{sec:theory_res}

An ideal procedure of proving that the observed changes in tumbling-state periods
$P_\phi$ and $P_\psi$ are due to the radiation (YORP) torques would require a
highly-reliable theoretical model (numerical propagation of 2012~TC4's rotation
state with appropriate torques included) employed to fit all available
observations (in our case data from October 2012 and 2017). Obviously, the
only ``comfort'' of this analysis would be to possibly adjust some
free (unknown) parameters. Unfortunately, such a plan is presently too ambitious,
and thus we resort to a simpler way. 

Recall the much easier situation when the YORP effect has been searched
(and detected) for asteroids rotating in the lowest-energy mode, namely about the
shortest principal axis of the inertia tensor. In this case, YORP results in
a secular change of the unique rotation period $P$ (as in $P_\phi$ and $P_\psi$,
when the body tumbles). The measurements are rarely precise enough, and the
effect strong enough, to directly reveal the change in the period $P$ 
\citep[see, though, an exception for 54509 YORP,][]{Low.ea:07}. More often, one
uses the fact that the linear-in-time change in $P$ produces a quadratic-in-time
effect in the rotation-phase. Linking properly the asteroid rotation phase
over many observation sessions with an empirical quadratic term helps to characterize
changes of $P$ which are individually too small to be determined from
one-apparition observations \citep[see, e.g.,][]{Vok.ea:15}. This approach
adopts an empirical magnitude of the quadratic term in the rotation phase and simply 
solves it as a free parameter (not combining it with a theoretical model at that 
stage). Interpretation in terms of the YORP effect is done only aposteriori, when
the fitted amplitude of the phase-quadratic term is compared with a prediction
from the YORP model. And even then, the comparison is often not simple,
because the model prediction for YORP is known to depend on unresolved
small-scale irregularities of the body shape. At several occasions one had
to satisfy with a factor of few difference accounting for the model inaccuracy.
\begin{figure}[tp]
\begin{center}
 \includegraphics[width=15cm]{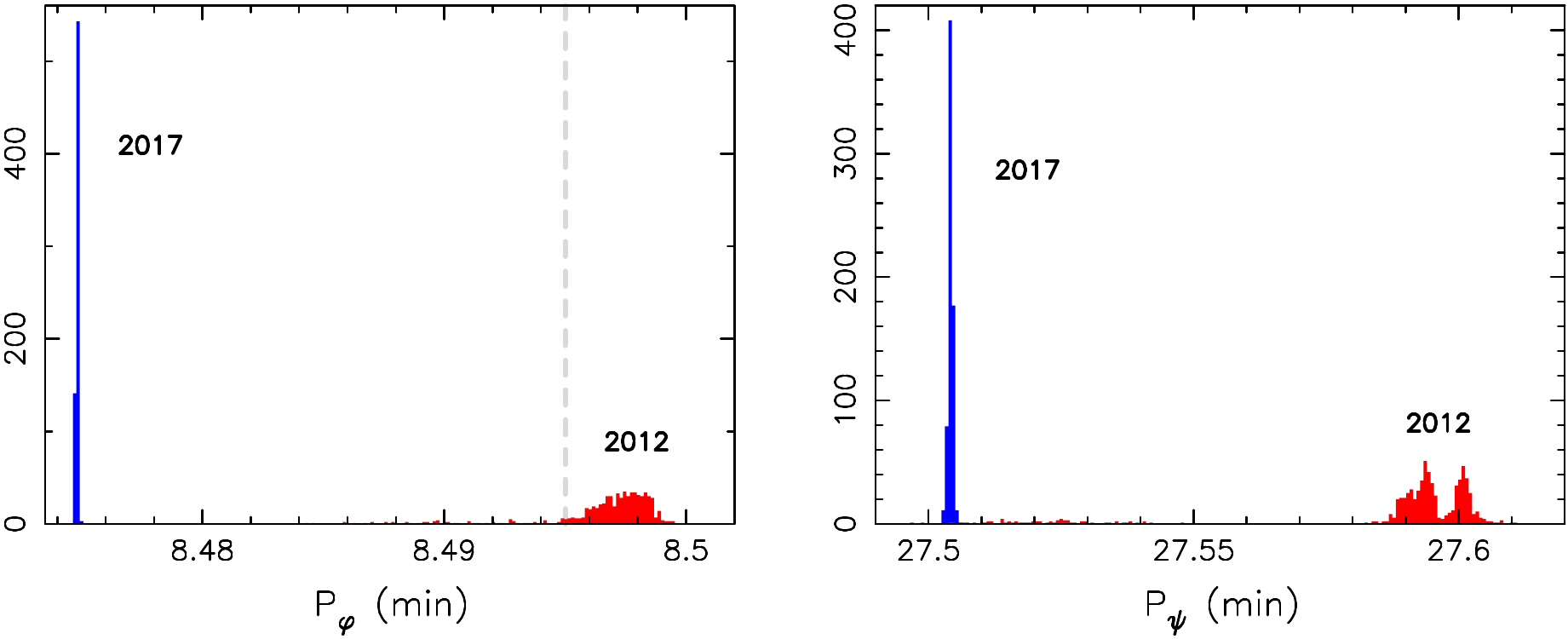}
\end{center}
\caption{Distribution of periods $P_\phi$ (left) and $P_\psi$ (right) for 687
 models of 2012~TC4 from our numerical simulation containing both gravitational
 and radiation (YORP) torques. The individual models sample possible initial
 orientation of the angular momentum vector $\mathbf{L}$ in the inertial space,
 orientation of the body-frame in the inertial space and
 slight shape variants of the body. All models were constructed using the October 2017
 observations and therefore they are referred to the common epoch MJD58032.19.
 Blue histograms are computed from these initial data and they are identical to
 those in Fig.~\ref{fig:histogram_periods}. The red histograms correspond to the rotation state of 2012~TC4 at MJD56209.88, nominal epoch of the October 2012 observations (the mean $P_\phi$ value from the observations is shown with the vertical dashed line). Unlike in
 Fig.~\ref{fig:histogram_periods}, the 2012 data here were computed from the spin
 state vectors obtained from our numerical propagation of the 2012~TC4 rotation starting
 in October 2017.}
\label{f4}
\end{figure}

Things are quite more complex when the body tumbles. First of all, it is not clear how
to set the empirical approach from above and apply it in this situation. In the
same time, direct modeling approach is probably even less accurate than in the
case of rotation about the principal axis of the inertia tensor. Not only
the worry about the role of unresolved small-scale irregularities remains, but the present
YORP model is restricted to the zero thermal conductivity limit (see, e.g.,
\cite{Vok.ea:07} for numerical approach, and \cite{Cic.Sch:10,
Bre.ea:11} for analytical studies). Yarkovsky acceleration for tumblers
was evaluated with thermal models \citep[e.g.,][]{Vok.ea:05, Vok.apophis:15}, 
but in these cases $P_\phi$ and $P_\psi$ were slightly
tweaked to make them resonant (an approach we cannot afford here).
In this situation, we adopted the following simple procedure. 

\subsubsection{Model based on 2017 data}
We start with a set of models based uniquely on the most reliable and accurate observations from the 2017 apparition.
In particular, we constructed
687 variants of the 2012~TC4 physical model (Sec.~\ref{sec:model_2017}). They are all
very similar, because they sample tight parameter variations
all resulting in acceptable fits of the data. These represent (i) slightly modified
initial rotation parameters (Euler angles and their derivatives, as well as
inertial space direction of the rotational angular momentum vector), and (ii)
slight shape variants of the body. The initial epoch MJD58032.19 was common to all
variant models. Using these initial data and shape models, we propagated all 687
clone realizations of 2012~TC4 backward in time to the epoch MJD56209.88,
characteristic of the October~2012 observations. We used our numerical approach
described above with both gravitational and radiation (YORP) torques included.
For the latter, we assumed an effective size $D=10$~m (corresponding to a sphere
of the same volume of the models) and bulk density $\rho=1.4$ g~cm$^{-3}$. The above-mentioned rescaling rule, namely invariance to $\rho\,D^2$, may allow us to transform our results to other combinations of $D$ and $\rho$ values.

The evolutionary tracks of angular momentum $L$ and energy $E$ secular changes due to
the YORP torques mostly resemble those from Fig.~\ref{f3}. For some model variants,
which were more different from the nominal one, the slope of the overall secular
change in $L$ and/or $E$ was shallower or steeper. At the initial and final epochs
of our simulations we determined $P_\phi$ and $P_\psi$ periods from a short numerical simulation of a free-top model. We also verified that the
$P_\psi$ values correspond exactly to those provided by the analytical formula
(\ref{e5}). Figure~\ref{f4} shows our results. The blue histograms are for the
initial data, i.e., the October 2017 rotation state. Because the observations
were numerous and of a good quality, the model variants differ only slightly and
the $P_\phi$ and $P_\psi$ distributions are tight (they also match
those from the Fig.~\ref{fig:histogram_periods}). The red histograms in Fig.~\ref{f4} were determined from
the last epoch of our numerical runs and correspond to the predicted rotation
state of 2012~TC4 in October~2012. We note both period appreciably changed, as
already anticipated from preliminary simulation shown in Fig.~\ref{f3}. These
$P_\phi$ and $P_\psi$ distributions are obviously less tight than the initial ones,
reflecting different evolutionary tracks of the individual clone variants. Our
prime interest is to compare the red distribution in Fig.~\ref{f4} (from simulations)
to the red distributions in Fig.~\ref{fig:histogram_periods} (from the 2012 observations, also highlighted withe dashed line for $P_\phi$ period). First,
we note that the match of the $P_\phi$ periods is surprisingly good. The mean value
$8.495$~min of the observations rather well corresponds to the mean value 
$8.497$~min of the simulated data, which have comfortably large dispersion of
$0.003$~min to overlap with the observed data; in fact, the shift in $P_\phi$ is even larger than required. Interestingly, the comparison
is not as good in the $P_\psi$ period. The model-predicted value $27.59\pm 0.02$~min
(formal uncertainty) is short to explain the observations which provide on average
$27.87$~min. Still, the model indicates a significant shift from the initial
value $27.5070\pm 0.0002$~min. Nevertheless, to reach the value from the 2012
observations the shift would need to be about $3.7$ times larger. We do not know
the reason for this difference, at all likelihood also related to the misbehaviour in the rotational energy evolution (see above). We suspect that the overly simple modeling
of the YORP effect, such as the surface thermal inertia and/or the unresolved small-scale shape irregularities, play
and important role here (note that a factor of $3$ between the observations and
model-prediction was also seen in the cases when YORP was detected for asteroids
rotating about the principal axis of the inertia tensor). The fact that some deeper aspects of the model are not characterized well, as witnessed by the opposite sign of the energy evolution, implies that our results cannot be
easily reconciled with the observations by a simple rescaling of size $D$ and
bulk density $\rho$. We have verified that the accumulated shifts in both
$P_\phi$ and $P_\psi$ periods are proportional to $\rho\,D^2$. Thus, the $P_\psi$
mismatch could be explained by assuming the 2012~TC4's size is $\simeq 5.2$~m,
but this would produce factor $\simeq 3.7$ inconsistency in $P_\phi$ period
(making the modeled value larger than observed). Instead, we believe that some
missing details of the YORP modeling, which result in different effects on 
$P_\phi$ and $P_\psi$, are responsible for the difference.

\subsubsection{Model based on a combination of 2012 and 2017 data}
For sake of
comparison, we also repeated our analysis using models based on a combination of the
observations in 2012 and 2017 (Sec.~\ref{sec:model_2app}). Obviously at each of these epochs we considered different
parameters of the rotation state, but now we enforced the same shape model is used for both 
data-sets. This solution gives us an opportunity to consider two sets of
initial conditions for our simulation, both in 2012 and 2017, and analyses
predictions in the complementary epoch (integrating the rotation model once
backward in time and once forward in time). Obviously, in all cases our model
includes the gravitational and radiation torques, $D=10$~m effective size and
$\rho = 1.4$ g~cm$^{-3}$ bulk density as before (needed for the evaluation of
the radiation torques).
\begin{figure}[tp]
\begin{center}
 \includegraphics[width=15cm]{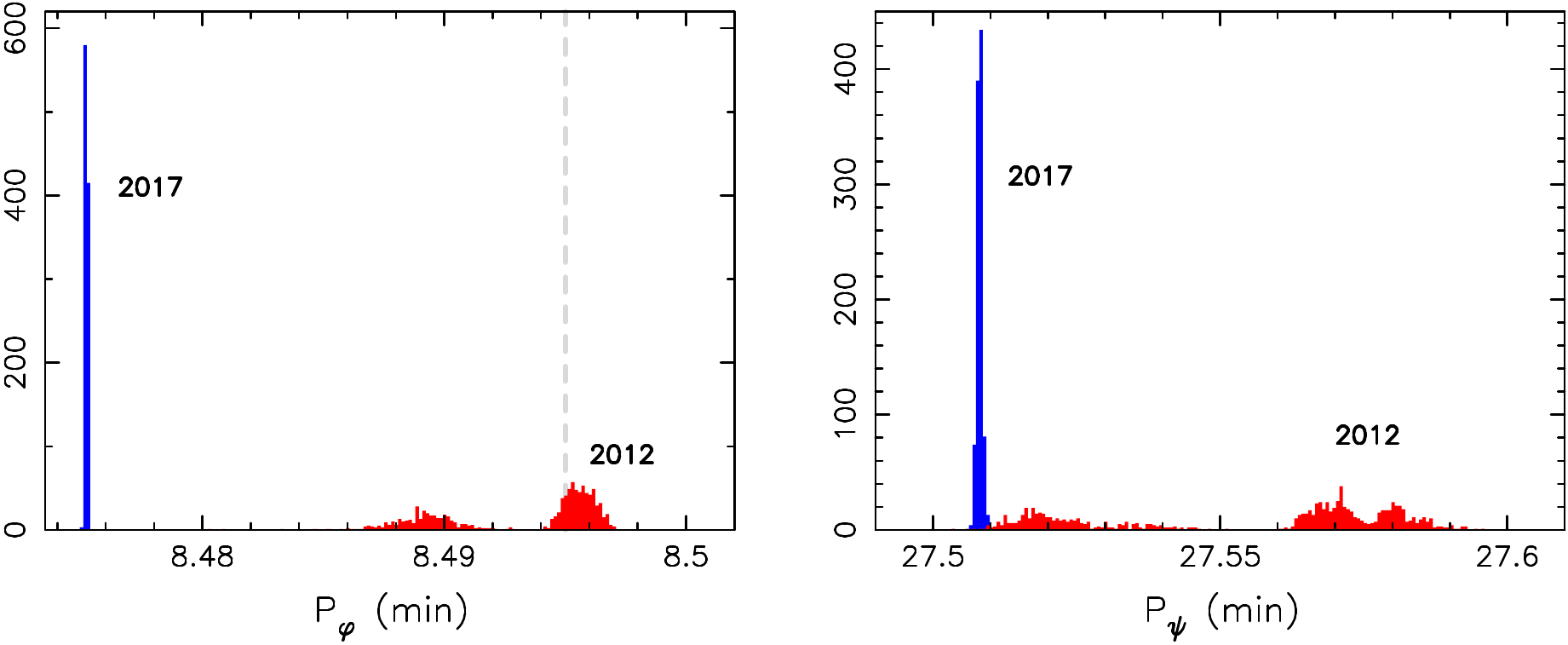}
\end{center}
\caption{The same as in Fig.~\ref{f4}, but now for nearly a thousand variant
 shape models of 2012~TC4 constructed from a combination of the 2012 and 2017 data.
 Data at the initial epoch MJD58032.19 (October 2017) are in blue. They have
 been propagated using our dynamical model to MJD56209.88 (October 2012), and
 the endstates of these runs served to compute $P_\phi$ and $P_\psi$ shown by
 the red histograms.}
\label{f5}
\end{figure}

We start with the case of the initial data in October 2017 and backward-in-time
integration. This is directly comparable with results above, when only observations
in 2017 were used. However, the models are slightly different in all their
aspects (initial conditions and shape), because now the 2012 observations play
the role in their construction. Results are shown in Fig.~\ref{f5}. While
slightly different then in Fig.~\ref{f4}, the principal outcome is the same:
(i) fairly satisfactory prediction for the $P_\phi$ period, while (ii) too small
change in the $P_\psi$ period. 
\begin{figure}[tp]
\begin{center}
 \includegraphics[width=15cm]{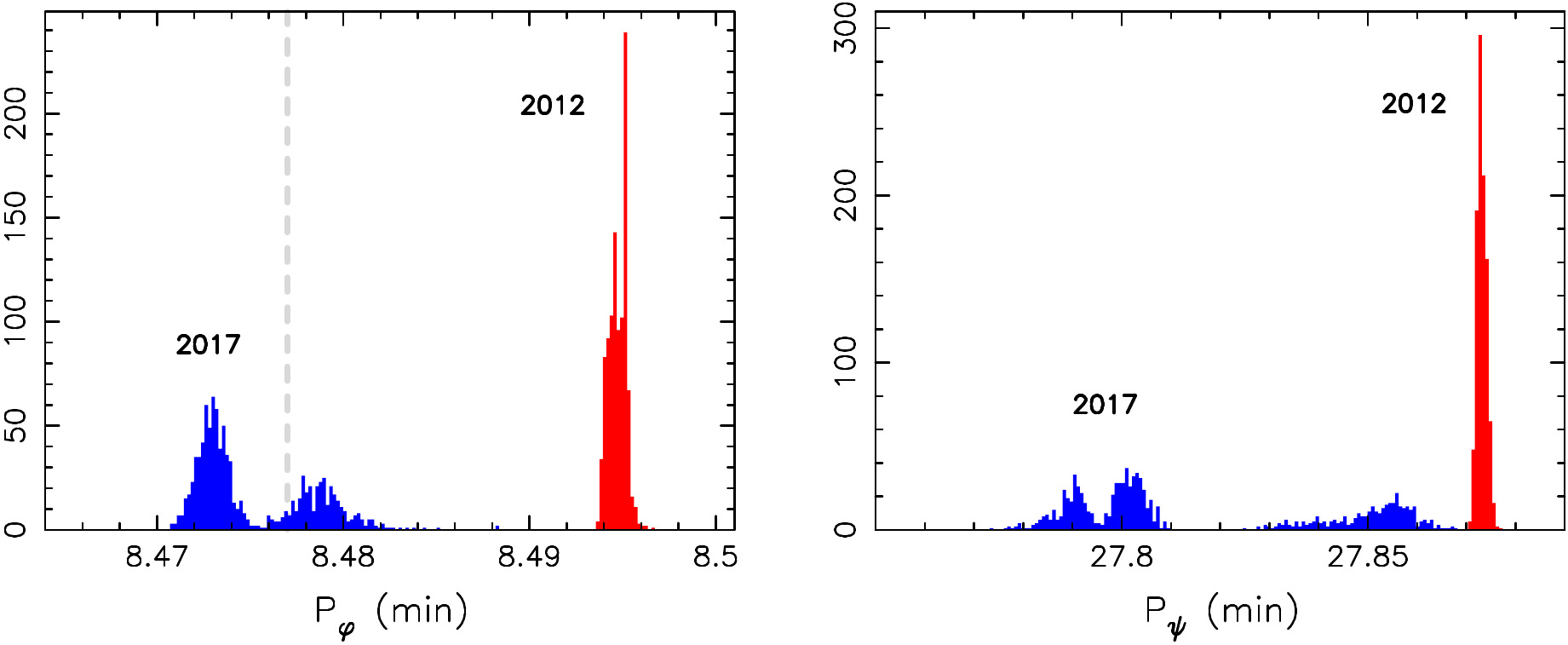}
\end{center}
\caption{The same as in Fig.~\ref{f5} (shape models constrained by both 2012 and
 2017 observations), but now propagation was performed from MJD56209.5 epoch
 in 2012 (red) to MJD58032.19 in 2017 (blue). The 2012 data constrain the rotation
 state solution less accurately and thus result in a larger scatter of the predicted
 periods $P_\phi$ and $P_\psi$ in 2017. The vertical dashed line shows mean value
 of the $P_\phi$ period from the October 2017 observations.}
\label{f6}
\end{figure}

We next consider the opposite case, namely initial condition in October 2012 and
model propagation forward in time to October 2017. Results are shown in Fig.~\ref{f6}.
Obviously, here the red histograms (from initial data in October 2012) are more
constrained than the blue histograms (resulting from rotation state vectors propagated
using our model to October 2017). The latter are more dispersed than the red histograms
in Fig.~\ref{f5}, because the less numerous and accurate observations in 2012 constrain
the models with lower accuracy. Nevertheless, the principal features of the solution
are still present: (i) the $P_\phi$ period changed adequately for majority of cases,
while (ii) the $P_\psi$ period changed too little.

\subsubsection{In summary}
So while we are not able to provide an exact proof that the
change in $P_\phi$ and $P_\psi$ periods between the 2012 and 2017 apparitions of 2012~TC4
are due to the YORP effect, we consider the difference between the observations and model
predictions can be accounted for the model inaccuracy.

\section{Discussion}
In the previous section, we demonstrated that the observed change of the rotation 
state of 2012~TC4 in between the two apparitions in 2012 and 2017 may be possibly
explained as a result of the YORP effect. We also showed that the effect of
the close encounter in October~12, 2012 on the rotation state was minimum, at
least in a rigid-body approximation. However, since the YORP model --for the reasons explained-- did
not provide an exact match of the observations, and even left unresolved the
issue of the observed energy change versus the model prediction, it is
both useful and necessary to also briefly analyse possible alternative
explanations. Here we discuss in some detail two plausible processes. We leave
aside a third possibility, notably a mass-loss from the surface
of 2014~TC4 sometime in the period between the two observation campaigns (or
during the 2012 Earth encounter). At the first sight, this may look an attractive
explanation because the fast rotation of this body implies formally negative
gravitational attraction at the surface. Therefore, it takes only the effect
of breaking cohesive bonds near the surface to make part of the body escape.
This event would have an influence on rotational energy and angular momentum,
both directly by the quanta carried away by escaping mass but also by a change
of the TC4's tensor of inertia. While possible, caveats of this model are twofold: (i) first, the observational data do not have resolution to provide a conclusive information about a shape change of the body (which may not be large for the effect to work at the one-per-mile level), and, (ii) more importantly, in most scenarios the process would lead to a decrease of the rotational angular momentum of TC4. Unlike in the two processes discussed below, we are not able to simply estimate likelihood of this process.

\subsection{Internal energy dissipation effects}
Individual solutions of
2012~TC4's rotation parameters in 2012 and 2017 indicate that periods $P_\psi$
and $P_\phi$ decreased in the latter epoch (Fig.~\ref{fig:histogram_periods}). In terms of osculating 
approximation with a free-top model it implies that the wobbling motion of the 
angular momentum vector ${\bf L}$ in the body-fixed frame moved toward the fundamental
mode of its direction along the shortest axis of the inertia tensor. Note the trajectory
of ${\bf L}$ in the body-fixed frame is uniquely parametrized with $p=2BE/L^2$
\citep[see, e.g., the Appendix~\ref{methods} and][]{Lan.Lif:69}. In quantitative terms, the
change from 2012 to 2017 state is expressed by $\delta p\simeq -6.2\times 10^{-3}$ (Table~\ref{tab:parameters}).
The average rate over $\delta t\simeq 5$~year interval would thus be $\delta p /
\delta t \simeq -3.9\times 10^{-11}$~s$^{-1}$.

In the YORP model presented above, the change in $p$ was a composition of changes
both in the energy $E$ and angular momentum $L$. In fact, both $E$ and $L$ increased 
from 2012 to 2017 (Fig.~\ref{f3}), but the composite effect was a decrease in $p$, in this case $\delta p\simeq -4.6\times 10^{-3}$, a similar value to that directly determined from osculating $L$ and $E$ above. Another
processes may lead to approximately the same results by producing a different
combination of energy and angular momentum changes. For instance, effects of
material inelasticity result directly in energy dissipation while preserving
angular momentum. In this case, both $E$ and $p$ decrease with a direct relation
$\delta E \simeq (L^2/2B)\,\delta p$. 

In order to explore whether the observed effect of the 2012~TC4's spin change
could even be plausibly matched by internal energy dissipation we used the model presented
by \cite{Bre.ea:12}. These authors assumed a fully triaxial geometry of
the body, but restricted their analysis of energy dissipation to the empirical
description with a quality factor $Q$ (see an alternative model of \cite{Fro.Efr:17}, 
where the authors describe the energy dissipation using a Maxwell
viscous liquid but allow only a biaxial shape of the body). With these assumptions, 
they expressed the secular (i.e., wobbling-cycle-averaged) rate of energy change
in the following form
\begin{equation}
 \frac{\delta E}{\delta t}\simeq - \frac{a^4 \rho m \Omega^5}{\mu Q}\,\Psi\; ,
   \label{diss}
\end{equation}
where $a$ is semimajor axis of the body's triaxial approximation, $\rho$ its
density, $m$ its mass, $\Omega = L/C$ and $\Psi$ a rather complicated factor
depending on nutation angle $\theta_{\rm nut}$ (i.e., tilt between ${\bf L}$ and
the shortest body axis; we find $\theta_{\rm nut}$ oscillates between $\simeq 16^\circ$
and $\simeq 46^\circ$, with a mean $\simeq 30^\circ$), body-axes ratios and Lame coefficients
\citep[see][Sec. 4.3]{Bre.ea:12}. Finally, $\mu$ is the Lam\'e shear modulus 
(rigidity) and $Q$ the quality factor, expressing empirically the energy 
dissipation per wobbling cycle. The product $\mu Q$ is characteristic to studies 
involving energy dissipation in planetary science since the pioneering work of
\cite{Bur.Saf:73} \citep[though, see already][]{Pre:58}. While highly 
uncertain, typical values of this parameter for asteroids range between $10^{11}$ 
and $5\times 10^{12}$~Pa \citep[e.g.,][]{Har:94}. Using $\delta E/\delta t
\simeq (L^2/2B)\,\delta p / \delta t$, with the above mentioned $\delta p / \delta t$
value, we can now use (\ref{diss}) to infer what values of $\mu Q$ would be needed
to explain the change in 2012~TC4's tumbling state in between 2012 and 2017
close approaches. The remaining unknown parameter is $\Psi$, which we
estimate to be $\simeq (1-5)\times 10^{-3}$. Plugging this value to (\ref{diss}),
we find $\mu Q$ ranging from $4\times 10^5$ to $4\times 10^6$~Pa. Remarkably, such
values are four to five orders of magnitude smaller than the usually adopted
estimates. Therefore, unless the energy dissipation by internal friction is
extraordinarily high (and thus the $\mu Q$ value very small), this process
cannot explain the observed rotation change of 2012~TC4. Note additionally, that we considered derivation of the needed $\mu Q$ value within the energy dissipation model as a useful exercise to match the energy change. Such a model, however, would not explain the observed angular momentum change.

\subsection{Impact by an interplanetary particle}
Another alternative process
to the radiation torques is that of an impact by interplanetary meteoroid. However,
in the following we provide an argument that the likelihood of this to happen at
the level needed to explain the 2012~TC4 data is again very small. To that end we
used information from \cite{Bot.ea:20}. This paper determined meteoroid flux
on a small asteroid (101955) Bennu using a state-of-art model MEM-3 allowing to
predict parameters of meteoroids impacting a target body orbiting between Mercury
and the asteroid belt \citep[e.g.,][]{Moo.ea:20}. Note that the orbit of Bennu is similar
to 2012~TC4 and we shall neglect the small flux differences that could result
from a small orbital dissimilarity of these two objects (if anything, the flux would
be slightly larger on Bennu, because of its orbit closer to the Sun). In their Fig.~2, 
\cite{Bot.ea:20} show that 5~mg interplanetary particles, approximately
$2$~mm in size, strike Bennu with frequency of $\simeq 60$ per year with a 
median impact velocity little less than $\simeq 30$~km/s.
For 2012~TC4 we only need to rescale this number to account for a much smaller
size: Bennu is a $\simeq 500$~m size asteroid, while a characteristic size
of 2012~TC4 is only $\simeq 10$~m. Therefore, the 5~mg flux on 2012~TC4 is
about $\simeq 2.4\times 10^{-2}$ per year. The chances to be hit by such a particle
in $5$ years is therefore merely $\simeq 0.12$. However, even if it
has happened, the dynamical effect would be minimum. Estimating the change in
rotational angular momentum $L$ plainly by $\delta L/L \simeq (m/M)(v_{\rm imp}/
R\omega)$, where $m$ and $M$ are masses of the particle and 2012~TC4, $v_{\rm imp}$ is
the impact velocity, and $R$ and $\omega$ are 2012~TC4's characteristic radius
and rotational frequency, we would obtain $\delta L/L \simeq 2\times 10^{-7}$.
At least ten thousand times larger effect would be needed to approach the level
observed for 2012~TC4, and this would require an impacting particle at least
twenty times larger, i.e. $5$~cm or more. Because MEM-3 incorporates flux
dependence on mass from \cite{Gru.ea:85}, thus $\propto m^{-4/3}$, the
chances that 2012~TC4 was hit by a $5$~cm meteoroid in between 2012 and 2017 is 
only $\simeq 6\times 10^{-7}$. We thus conclude that chances that the observed 
effect was produced by an impact of interplanetary particle is negligibly
small and may be discarded.

\section{Conclusions}

Photometric data of 2012~TC4 collected during its two close approaches to the Earth in 2012 and 2017 clearly show that this asteroid is in an excited rotation state. Fourier analysis of 2012 and 2017 data sets finds two unique periods in the signal and these two periods are significantly different, which means that the rotation state of TC4 must have changed slowly between 2012 and 2017 or suddenly during the 2012 flyby (the 2017 flyby was after photometric observations). This detection of period change is robust and not model-dependent.

The periods detected by Fourier analysis were used to constrain physical rotation and precession periods of the tumbling rotation state. We found only one physically acceptable solution that fits photometric data, namely free-tumbling situation about the shortest axis of the inertia tensor (SAM mode). When modeling light curve sets from 2012 and 2017 separately, the shape models are similar with about the same direction of angular momentum vector but the rotation and precession periods are significantly different and there is no combination of parameters that would provide an acceptable fit to the whole data set. The change of physical periods of tumbling is consistent with the change revealed by Fourier analysis. Including this period change into our model, we were able to fit all available photometry from both apparitions. The difference in periods for 2012 and 2017 apparitions is much larger than any possible random or model error, so this model-based detection of periods change is significant and robust.

Having detected the period change and creating a physical model of TC4, we looked for a possible explanation for this change. First, we show that the effect of close encounter in 2012 on the rotation state was negligibly small compared to the detected change of the rotation state. Second, we show that a plausible explanation is the YORP effect -- the numerical simulation of the rotation dynamics based on our shape model of TC4 gives a general agreement with observed periods change. Although the match is not ideal, we believe that the discrepancy is caused by simplification in our YORP model and uncertainties in the shape model and other parameters. We also show that the other two possible mechanisms that could affect the rotation state -- namely the internal energy dissipation and impacts of interplanetary particles -- are too small to cause the measured effect, so YORP remains the only plausible explanation of the observed change of the rotation state of 2012~TC4. Accepting this explanation, this is the first detection of YORP acting on a tumbling asteroid.

\acknowledgements
  This research is supported by Korea Astronomy and Space Science Institute (KASI). 
  The work at Charles University and Ond\v{r}ejov Observatory was supported by the Czech Science Foundation (grant 20-04431S).
  The work by C.-H. Kim was supported by the Basic Science Research Program through the National Research Foundation of Korea funded by the Ministry of Education (2018R1D1A1A09081827, 2020R1A4A2002885).
  This research has made use of the KMTNet system operated by the Korea Astronomy and Space Science Institute (KASI) and the data were obtained at SAAO in South Africa. 
  We appreciate all astronomers who uploaded or shared their previously published light curves of 2012 TC4 used in this paper.

\appendix

\section{Rotational dynamics of 2012~TC4} \label{methods}
In this Appendix, we summarize variables and mathematical approach
used for propagation of 2012~TC4 rotation state in between the 2012 and 2017
epochs (Section~\ref{sec:theory}). This is obviously a classical piece of mechanics which can be found
in many textbooks \citep[e.g.,][]{Lan.Lif:69, Gol:80}. For that 
reason we keep our description to a very minimum.

The kinematical part of the problem describes orientation of the
asteroid in the inertial frame. For simplicity we assume the asteroid is a rigid
body, allowing us to define unambiguously a proper body-fixed frame. The
easiest choice has (i) the origin in the asteroid's center-of-mass, and (ii)
the axes coinciding with the principal axes of the inertia tensor $\mathbf{I}$
(therefore $\mathbf{I}={\rm diag}(A,B,C)$, with $A\leq B\leq C$). Transformation 
between the inertial and body-fixed frames is conventionally parametrized by a set 
of Euler angles, most often the 3-1-3 sequence of the precession angle $\phi$, 
the nutation angle $\theta$, and the angle of proper rotation $\psi$. However,
instead of the three Euler angles $(\phi,\theta,\psi)$ we use here the
Rodrigues-Hamilton parameters $\mbox{\boldmath$\lambda$}=(\lambda_0,\lambda_1,
\lambda_2,\lambda_3)$ \citep[e.g.,][]{Whi:1917}. Their relation to the Euler angles
is given by: (i) $\lambda_0+\imath \lambda_3 = \cos\frac{\theta}{2}\,\exp\left[
\frac{\imath}{2} \left(\psi+\phi\right)\right]$, and (ii) $\lambda_2+\imath \lambda_1
= \imath \sin\frac{\theta}{2}\,\exp\left[\frac{\imath}{2} \left(\psi-\phi\right)\right]$
($\imath$ is a complex unit). One can easily verify a constraint: $\lambda_0^2+
\lambda_1^2+\lambda_2^2+\lambda_3^2=1$ (in our numerical runs satisfied with a 
$\leq 10^{-13}$ accuracy). The sacrifice of using four instead of three parameters
pays off in at least two advantages. First, parametrization by Euler angles is 
unstable when, or near, $\sin\theta=0$ state. No such problem occurs when using 
the Rodrigues-Hamilton parameters which provide a uniformly non-singular description
of the rotation. Second, Euler-angle parametrization necessarily 
requires use of trigonometric functions. Instead, manipulation with the 
Rodrigues-Hamilton parameters is limited to simple algebraic functions, in fact
quadratic at maximum, as shown below in Eqs.~(\ref{e1}) and (\ref{e2}). For that 
reason the use of the four Rodrigues-Hamilton parameters does not even slow down 
the computations in a noticeable way. 

The rotation matrix $\mathbf{A}$ needed for the vector transformation from the
inertial frame to the body-fixed frame is a simple quadratic form of 
$\mbox{\boldmath$\lambda$}$, namely 
\begin{equation}
 \mathbf{A} = \left(
 \begin{array}{ccccc}
 \lambda_0^2+\lambda_1^2-\lambda_2^2-\lambda_3^2 & , & 2\left(\lambda_0\lambda_3+
  \lambda_1\lambda_2\right) & , & 2\left(\lambda_1\lambda_3-\lambda_0\lambda_2\right) \\
 2\left(\lambda_1\lambda_2-\lambda_0\lambda_3\right) & , & \lambda_0^2+\lambda_2^2-
  \lambda_1^2-\lambda_3^2 & , & 2\left(\lambda_0\lambda_1+ \lambda_2\lambda_3\right) \\
 2\left(\lambda_0\lambda_2+ \lambda_1\lambda_3\right) & , & 2\left(\lambda_2\lambda_3-
  \lambda_0\lambda_1\right) & , & \lambda_0^2+\lambda_3^2-\lambda_1^2-\lambda_2^2 \\
 \end{array}
 \right) \; . \label{e1}
\end{equation}
The inverse transformation is represented by a transposed matrix $\mathbf{A}^{\rm T}$.
Asteroid's rotation is represented with the angular velocity vector $\mbox{\boldmath$
\omega$}$, whose components in the body-fixed frame are $(\omega_1,\omega_2,\omega_3)$.
Their relation to the time derivatives of the Rodrigues-Hamilton parameters is
simply
\begin{equation}
 \frac{d \mbox{\boldmath$\lambda$}}{dt} = \frac{1}{2}\,\mathbf{P}\cdot 
  \mbox{\boldmath$\lambda$} \; , \label{e2}
\end{equation}
where
\begin{equation}
 \mathbf{P} = \left(
 \begin{array}{ccccccc}
  0 & , & -\omega_1 & , & -\omega_2 & , & -\omega_3 \\
   \omega_1 & , & 0 & , & \phantom{-}\omega_3 & , & -\omega_2 \\
   \omega_2 & , & -\omega_3 & , & 0 & , & \phantom{-}\omega_1 \\
   \omega_3 & , & \phantom{-}\omega_2 & , & -\omega_1 & , & 0 \\
 \end{array}
 \right) \; . \label{e3}
\end{equation}
This explicitly linear differential equation for $\mbox{\boldmath$\lambda$}$ cannot
be solved in a trivial way, because $\mathbf{P}=\mathbf{P}\left(\mbox{\boldmath$
\omega$}\right)$, and the angular momentum vector is a time-dependent variable.
The antisymmetry of $\mathbf{P}$ implies conservation of the above mentioned
quadratic constraint of $\mbox{\boldmath$\lambda$}$.

The dynamical part of the problem expresses the Newton's principle that a change
of the rotational (intrinsic) angular momentum $\mathbf{L} = \mathbf{I}\cdot\mbox{
\boldmath$\omega$}$ is given by the applied torque $\mathbf{M}$. Tradition has it to 
state this rule in the body-fixed frame, where $\mathbf{I}$ is constant and even
diagonal in our choice of axes, such that
\begin{equation}
 \frac{d\mathbf{L}}{dt} + \mbox{\boldmath$\omega$}\times \mathbf{L} = \mathbf{M}
  \; . \label{e4}
\end{equation}
Equations (\ref{e2}) and (\ref{e4}) define the problem of asteroid's rotation in our 
set of seven parameters $(\mbox{\boldmath$\lambda$},\mbox{\boldmath$\omega$})$. Once 
the torques $\mathbf{M}$ are specified, we numerically integrate this system of 
differential equations with the initial data determined from the set of observations (either forward in time if the 2012 data are used, or backward in time if the 2017 data are used). We use Burlish-Stoer integration scheme with tightly controlled accuracy.
We also note another useful quantity, namely the energy of rotational motion about 
the center given by $E=\frac{1}{2}\,\mbox{\boldmath$\omega$}\cdot \mathbf{L}$. In a
classical problem of a free top (i.e., $\mathbf{M}=0$), both $E$ and $\mathbf{L}$
in the inertial frame are conserved. In the body-fixed frame only $L=|\mathbf{L}|$
is constant. Nevertheless, conservation of $E$ and $L$ (together with the principal values of the inertia tensor $A$, $B$ and $C$) uniquely determines the
wobbling trajectory of $\mathbf{L}$ in the body-fixed frame \citep[e.g.,][]{Lan.Lif:69, Dep.Eli:93}. 
There are two options for this motion: (i) 
short-axis mode (SAM),
when $\mathbf{L}$ circulates about $+z$ or $-z$ body axis, or (ii) long-axis (LAM),
mode $\mathbf{L}$ circulates about $+x$ or $-x$ body axis. A useful discriminator
of the two is yet another conserved and non-dimensional quantity in the free-top 
problem, namely $p=2BE/L^2$: (i) SAM is characterized by $p$ values in between
$\beta=B/C$ and $1$, while (ii) LAM is characterized by $p$ values in between $1$
and $\alpha=B/A$. Note that $\Delta=B/p$ plays an important role in description
of the free-top problem using Hamiltonian tools \citep[e.g.,][]{Dep.Eli:93,
Bre.ea:11}. The free-top motion of $\mathbf{L}$ in the body fixed frame
is easily integrable using Jacobi elliptic functions. When plugged in the
kinematical equations (\ref{e3}), one can also obtain solution for $\mbox{\boldmath$
\lambda$}$ or the Euler angles $(\phi,\theta,\psi)$ \citep[e.g.,][]{Lan.Lif:69}. 
Those of $\psi$ and $\theta$ are strictly periodic with a period (SAM mode relevant 
for 2012~TC4 assumed here)
\begin{equation}
 P_\psi = \frac{C}{L}\, \frac{4\beta K\left(k\right)}{\sqrt{\left(1-\beta\right)
  \left(\beta/\alpha-p\right)}}\; , \label{e5}
\end{equation}
where $K(k)$ is complete elliptic integral of the first kind with the modulus $k$ 
given by
\begin{equation}
 k^2 = \frac{\left(\beta/\alpha-1\right)\left(p-\beta\right)}{\left(1-\beta\right)
  \left(\beta/\alpha-p\right)}  \label{e6}
\end{equation}
(the motion of $\theta$ has a periodicity $P_\psi/2$). The motion of the precession
angle $\phi$ is not periodic. Nevertheless a fully analytical solution still exists
and it is composed of two parts, the first of which has periodicity $P_\psi$ and a
second has another periodicity, generally incommensurable with $P_\psi$ \citep[e.g.,][]{Lan.Lif:69}. 
Yet it is both practical and conventional to define an approximate periodicity 
$P_\phi$ of $\phi$. We use definition of \citet{Kaa:01}, Eq.~(A.11), namely numerically determining an advance in $\phi$ angle over $P_\psi$ period (this, in principle, averages out contribution of the $P_\psi$-periodic part in the $\phi$ solution). When weak torques are applied, 
the free-top solution represents still a very useful (osculating) template with all
above-discussed variables such as $E$, $L$, $p$, $P_\psi$ or $P_\phi$ adiabatically
changing in time.

Finally, we discuss the torques used in our analysis. The first class is due to
gravitational tidal fields of the Sun and the Earth. Assume a point mass source
$M$ specified in the body-fixed frame of the asteroid with a position vector
$\mathbf{R}$. Using the quadrupole part of the exterior perturber tidal field
we have \citep[e.g.,][]{Fit:70, Tak.ea:13}
\begin{equation}
 \mathbf{M}_{\rm grav} = \frac{3GM}{R^5}\,\mathbf{R}\times \left(\mathbf{I}
  \cdot\mathbf{R} \right)\; . \label{e7}
\end{equation}
We neglect the formally dipole part of the tidal field, which could only
occur if the true center-of-mass of the asteroid is slightly displaced from the
assumed location \citep[determined by using the assumption of homogeneous density;
see, e.g.,][]{Tak.ea:13}. Note the position of all bodies, the
asteroid, the Sun and the Earth, are primarily determined using the numerical
integration of the orbital problem in the inertial frame (or, actually, displaced
heliocentric frame). In our case, we numerically integrated planetary orbits,
including the Earth, and 2012~TC4 in the heliocentric system by taking initial
data from NEODyS website (\url{https://newton.spacedys.com/neodys/}). We output the
necessary positions every $170$~s, enough for the purpose of the 2012~TC4's 
rotation dynamics. We also compared our solution with that available at the JPL
Horizons system (\url{http://ssd.jpl.nasa.gov/?horizons}), and found a very good
correspondence with tiny differences, not meaningful for our application.
The relative position $\mathbf{R}$ in (\ref{e7}) is determined by (i) the 
difference of the corresponding bodies in our orbital solution, and (ii) 
transformation to the body-fixed
frame. As a result $\mathbf{M}_{\rm grav} = \mathbf{M}_{\rm grav}\left(t,
\mbox{\boldmath$\lambda$}\right)$. The steep dependence $M_{\rm grav}\propto
R^{-3}$ implies that the Earth effect is non-negligible only during the close
encounters to this planet (e.g., Fig.~\ref{f1}).

Our analysis also includes the radiation torque known as the YORP effect 
\citep[e.g.,][]{Bot.ea:06, Vok.ea:15}. Because of the tumbling
rotation state of 2012~TC4, we resort to the simplest variant, namely a limit
of zero thermal inertia \citep[the effects of finite thermal inertia were studied
only for objects rotating about the shortest axis of the inertia tensor so far;
e.g.,][]{Cap.Vok:04, Gol.Kru:12}. In this
approximation the radiation torque is given by \citep[e.g.,][]{Rub:00, Vok.Cap:02}
\begin{equation}
 \mathbf{M}_{\rm YORP}= -\frac{2 F}{3c}\int H\left[\mathbf{n}\cdot\mathbf{n}_0\right]
  \left(\mathbf{n}\cdot\mathbf{n}_0\right)\,\mathbf{r}\times d\mathbf{S}\; , \label{e8}
\end{equation}
where $F$ is the solar radiation flux at the location of the asteroid and $c$
is the light velocity. The integral in (\ref{e8}) is performed over the surface
of the body characterized by an ensemble of outward-oriented surface elements
$d\mathbf{S}=\mathbf{n}\,dS$, $\mathbf{n}$ is the normal to the surface and
$\mathbf{r}$ is the position of the surface element with respect to the origin
of the body-fixed frame. The unit vector of the solar position in the body fixed
frame is denoted with $\mathbf{n}_0$, and $H[x]$ is the Heaviside step function
(its presence in the integrand of (\ref{e8}) implies that a non-zero contribution
to the radiation torque is provided by surface units for which the Sun is
above local horizon). In fact, our code includes even more complex feature
of self-shadowing of surface units, but this is not active in the case of
2012~TC4 whose resolved shape is convex. The factor $2/3$ on the right hand
side of Eq.~(\ref{e8}) is due to the assumption of Lambertian reradiation
from the surface. More complicated assumptions about directionality of the thermally emitted radiation from the surface, such as the beaming effects \citep[e.g.,][]{Roz.Gre:12}, are presently not implemented in the code. The lightcurve inversion obviously allows only a finite
accuracy in shape determination of the body, typically a convex polyhedron
with little more than thousand surface facets. It is known that already this fact is an obstacle to an accurate evaluation of the YORP torque, which may sensitively depend on smaller-scale surface irregularities, not resolved by our shape model. The formal integration in
Eq.~(\ref{e8}) is therefore represented with a summation over the surface units
of the resolved shape model. We use algebra from \cite{Dob:96} to
determine all necessary variables. This also means we assume constant density
distribution in the body.

\section{Light curve fits}
\label{sec:lc_fits}

    Here we show how the model fits the data. In all plots below, blue dots are individual photometric measurements, the red light curve is what the best-fit model form Sect.~\ref{sec:model_2app} predicts. The relative brightness on the vertical axis is scaled to have the mean value of one. Ticks on the horizontal axis are 10\,min apart and the scale is the same for all plots. More information about light curves can be found in Table~\ref{table:observation}.
    
  \begin{figure*}
   \centering
   \includegraphics[height=\textheight]{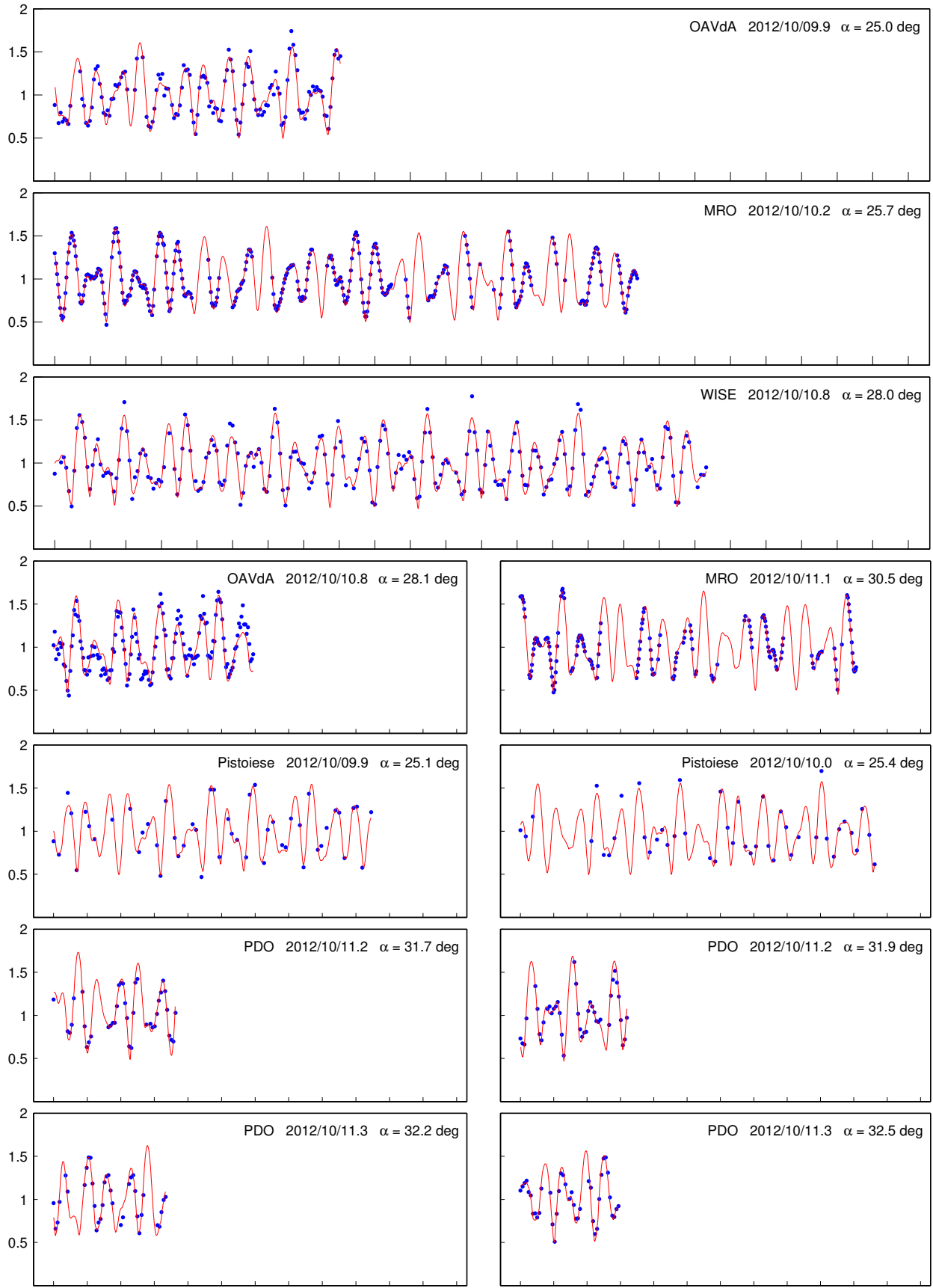}
   \caption{Light curve fits for data in 2012. The photometric light curve data shown in this figure is available as "Data behind the Figure."}
   \label{fig:lc_fit_2012_1}
  \end{figure*}

  \begin{figure*}[p]
   \centering
   \includegraphics[width=\textwidth]{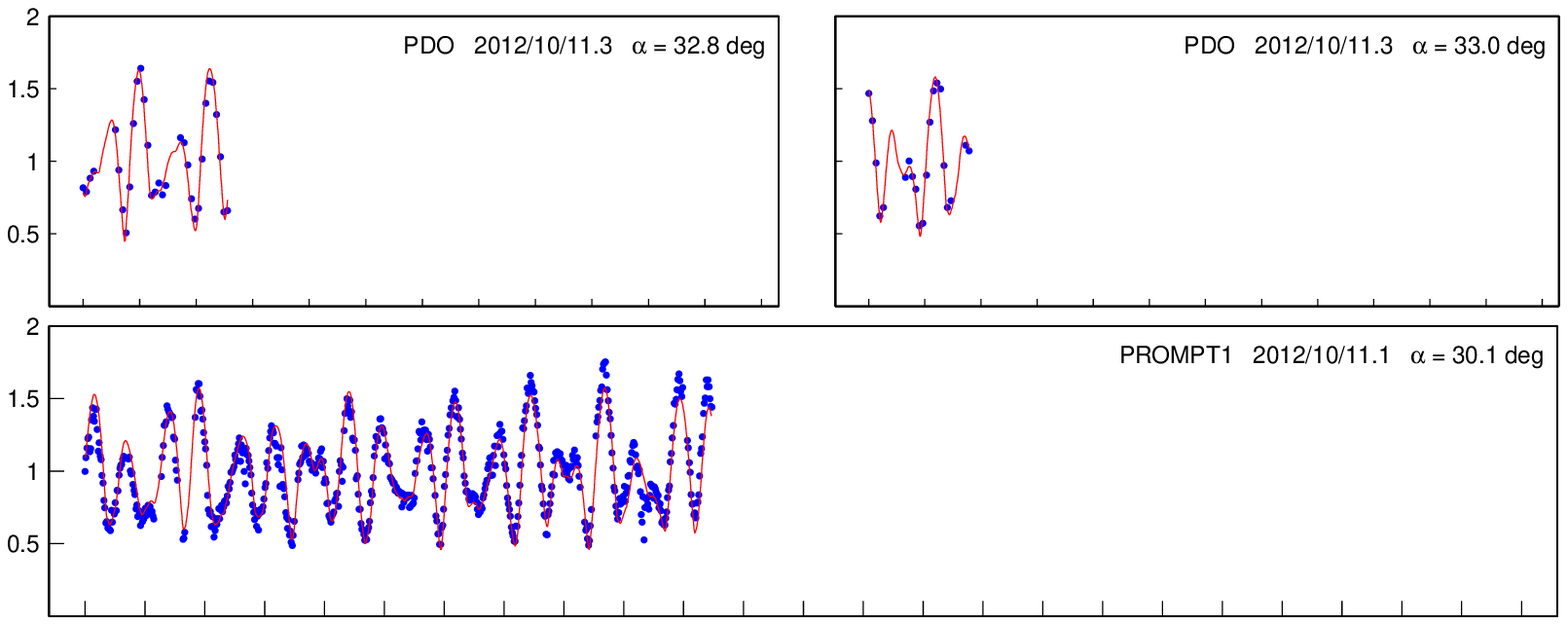}
   \caption{Light curve fits for data in 2012. The photometric light curve data shown in this figure is available as "Data behind the Figure."}
   \label{fig:lc_fit_2012_2}
  \end{figure*}

  \begin{figure*}[p]
   \centering
   \includegraphics[height=\textheight]{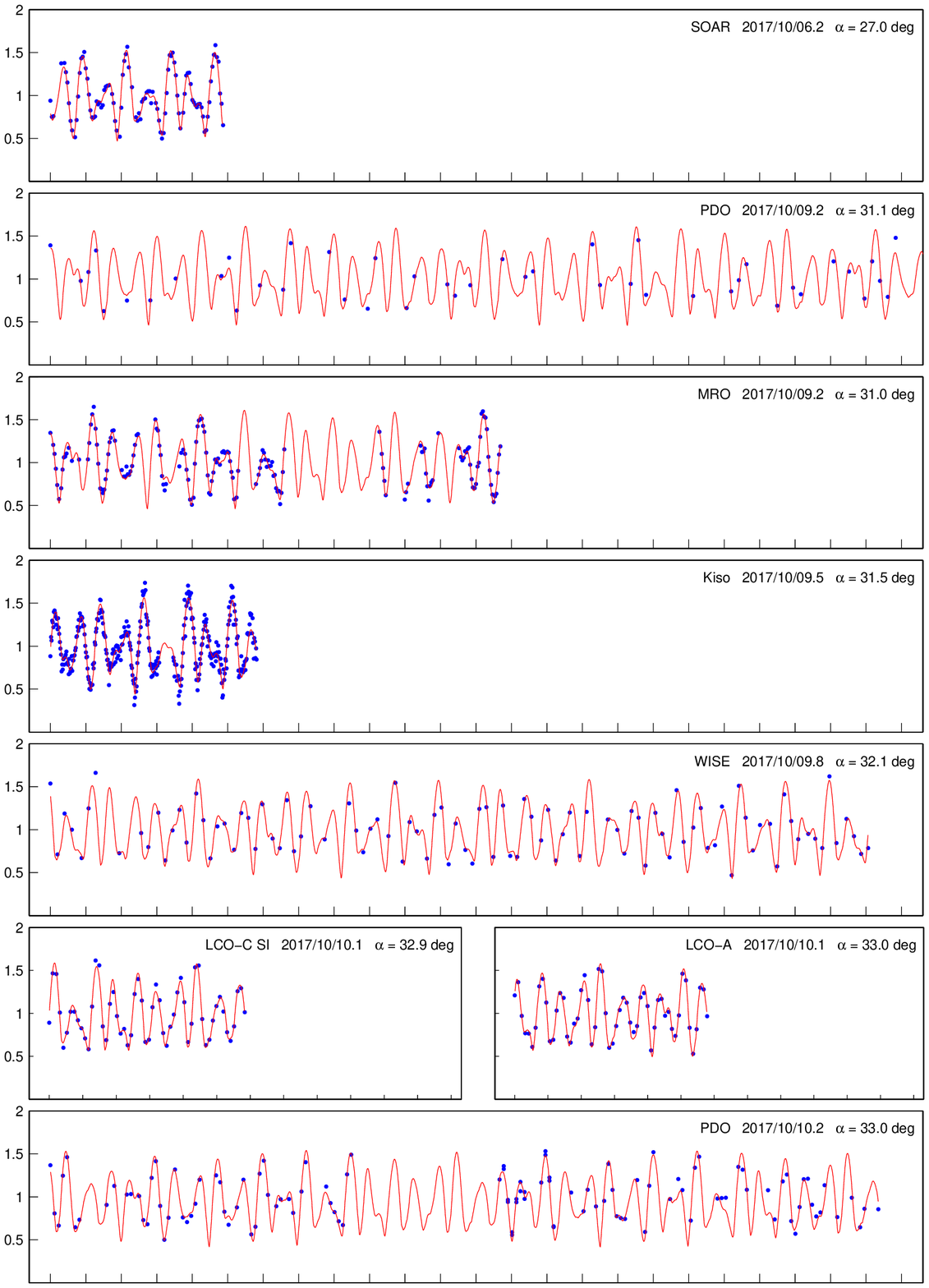}
   \caption{Light curve fits for data in 2017. The photometric light curve data shown in this figure is available as "Data behind the Figure."}
   \label{fig:lc_fit_2017_1}
  \end{figure*}

  \begin{figure*}[p]
   \centering
   \includegraphics[height=\textheight]{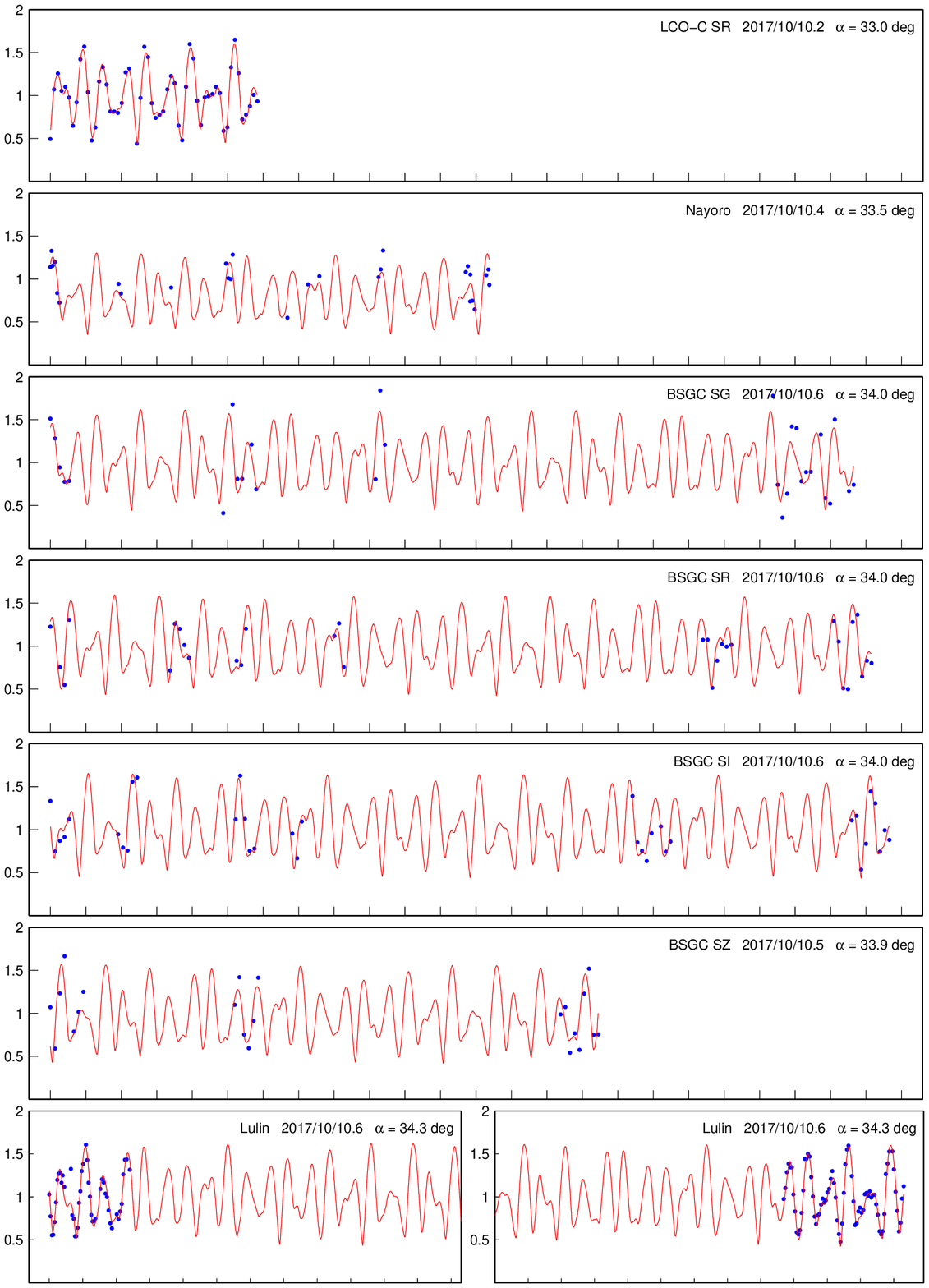}
   \caption{Light curve fits for data in 2017. The photometric light curve data shown in this figure is available as "Data behind the Figure."}
   \label{fig:lc_fit_2017_2}
  \end{figure*}

  \begin{figure*}[p]
   \centering
   \includegraphics[height=\textheight]{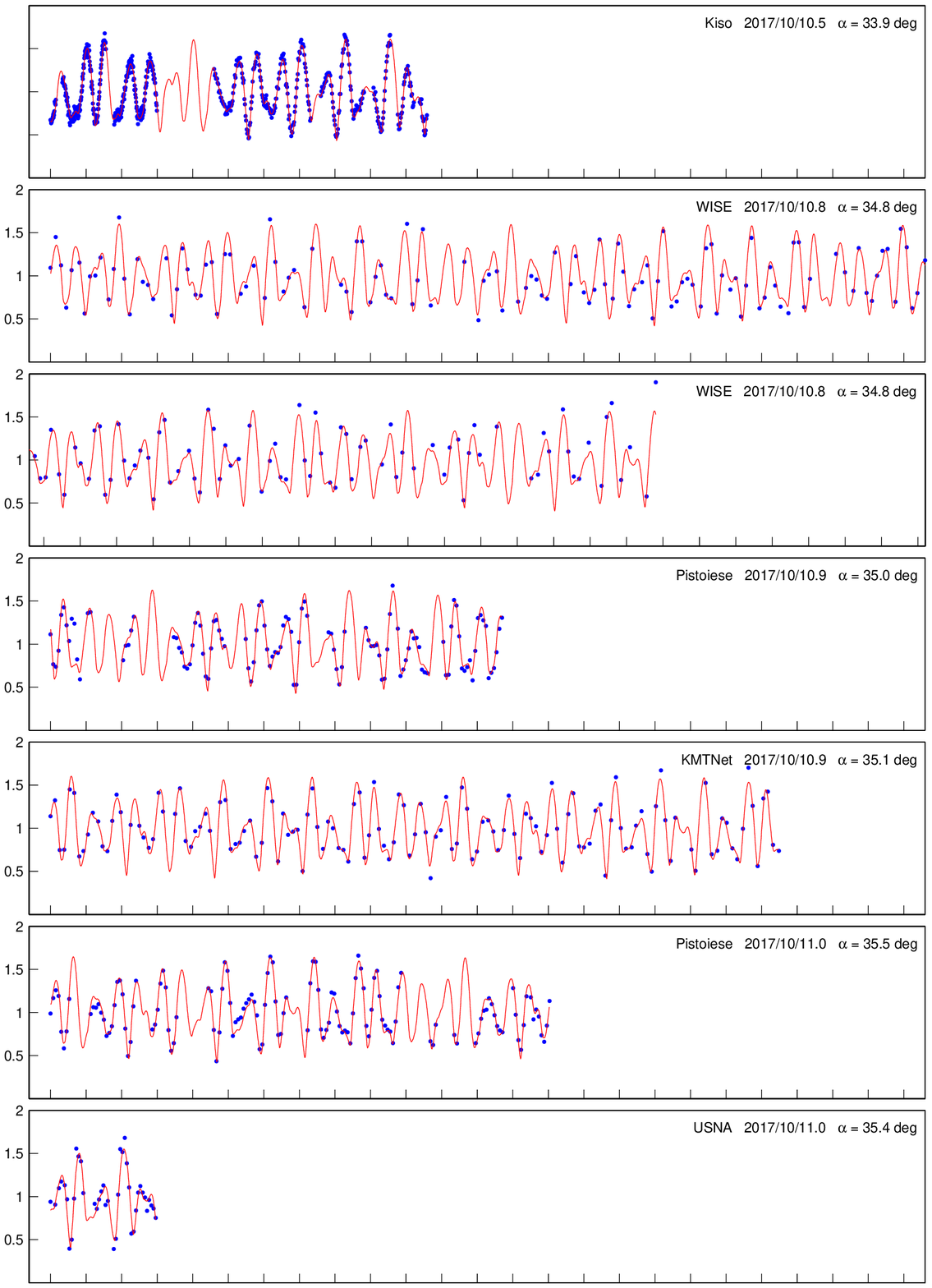}
   \caption{Light curve fits for data in 2017. The photometric light curve data shown in this figure is available as "Data behind the Figure."}
   \label{fig:lc_fit_2017_3}
  \end{figure*}

  \begin{figure*}[p]
   \centering
   \includegraphics[height=\textheight]{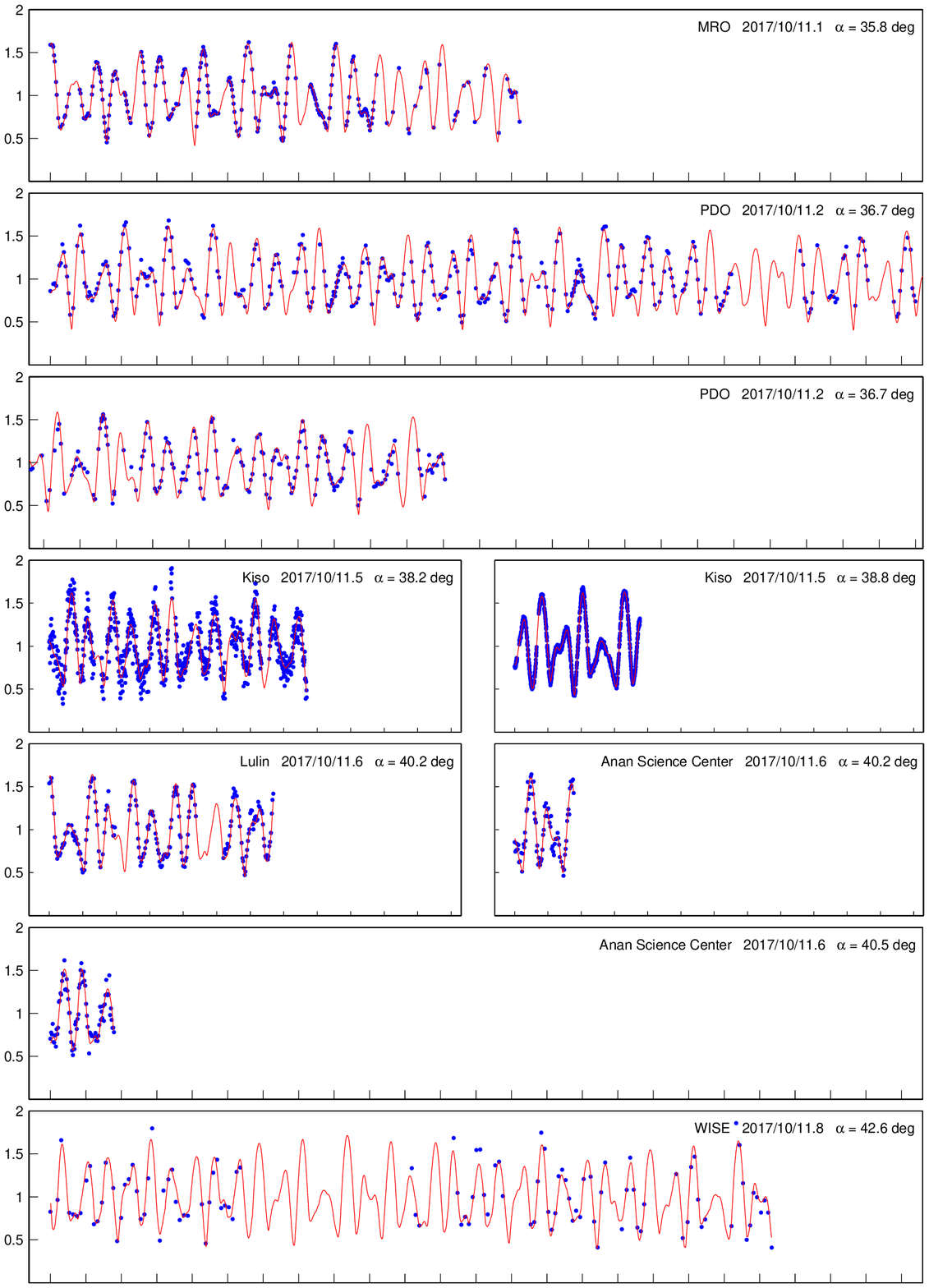}
   \caption{Light curve fits for data in 2017. The photometric light curve data shown in this figure is available as "Data behind the Figure."}
   \label{fig:lc_fit_2017_4}
  \end{figure*}

  \begin{figure*}[p]
   \centering
   \includegraphics[width=\textwidth]{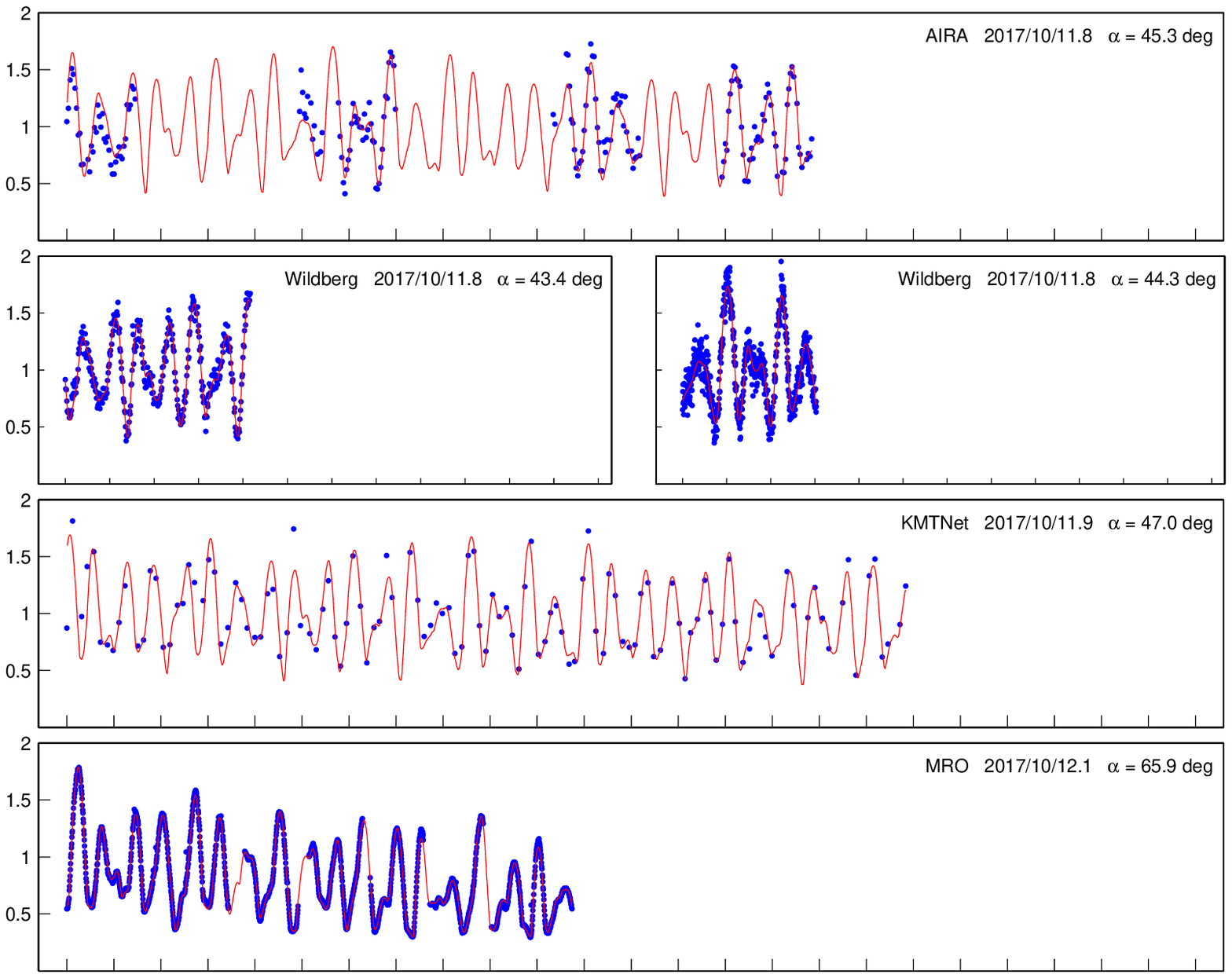}
   \caption{Light curve fits for data in 2017. The photometric light curve data shown in this figure is available as "Data behind the Figure."}
   \label{fig:lc_fit_2017_5}
  \end{figure*}

\clearpage

\bibliography{sample63,bibliography_all}{}
\bibliographystyle{aasjournal}

\end{document}